\documentstyle[epsf]{mn}
\def\gsim{\;\lower.6ex\hbox{$\sim$}\kern-7.75pt\raise.65ex\hbox{$>$}\;}
\def\lsim{\;\lower.6ex\hbox{$\sim$}\kern-7.75pt\raise.65ex\hbox{$<$}\;}

\title[Photometry and metal abundances of Hipparcos metal-poor stars ]
{Homogeneous photometry and metal abundances for a large sample of Hipparcos 
metal-poor stars \thanks{ Based on
data from the Hipparcos astrometry satellite and from Asiago and McDonald
Observatories}} 

\author[Clementini et al.]{G. Clementini$^1$, R.G. Gratton$^2$, E. Carretta$^2$,
       C. Sneden$^3$ \\
 $^1$ Osservatorio Astronomico di Bologna, Italy, 
      e-mail gisella@astbo3.bo.astro.it \\
 $^2$ Osservatorio Astronomico di Padova, Italy, 
      e-mail gratton@pd.astro.it, carretta@pd.astro.it \\
 $^3$ Department of Astronomy, The University of Texas at Austin, 
      e-mail chris@verdi.as.utexas.edu }
\date{}

\begin{document}
\maketitle

\begin{abstract}
Homogeneous photometric data (Johnson $V$, $B-V$, $V-K$, Cousins $V-I$ and
Str\"omgren $b-y$), radial velocities, and abundances of Fe, O, 
Mg, Si, Ca, Ti, Cr and Ni are presented for 99
stars with high precision parallaxes measured by the HIPPARCOS satellite.
These data have been previously used to assist the derivation of accurate 
distances and ages of galactic globular clusters.
Magnitudes and colours for the programme stars were obtained combining
and standardizing carefully selected literature data available in the 
Simbad data-base and $V$ and $B-V$ values measured by the Hipparcos/Tycho 
mission. Comparison of colours for our targets suggests: 
(i) ground-based and Tycho $B-V$'s agree well for colours bluer than 
0.75~mag, but have a lot of scatter for redder colours;
(ii) the Hipparcos $V-I$ colours have a very large scatter and a zero 
point offset of +0.02~mag compared to the literature values.

The programme stars have metal abundances in the range $-2.5<$[Fe/H]$<0.2$. The
spectroscopic observational data set consists of high dispersion 
($15\,000<R<60\,000$), high
$S/N$\ ($>200$) spectra obtained at the Asiago and McDonald Observatories 
for 66 stars. The
analysis is carried out following the same precepts used 
in previous papers of this series and includes corrections for departures
from LTE in the formation of the O lines. The main results are: 
(i) the equilibrium of ionization of Fe is well satisfied in
late F-early K-dwarfs; (ii) Oxygen and $\alpha-$elements are overabundant 
by $\sim 0.3$~dex.

This large homogeneous abundance data set has been used to 
recalibrate the abundance
scales of Schuster \& Nissen (1989), Carney et al. (1994), and Ryan \& Norris
(1991).
\end{abstract}


\section{Introduction}

The HIPPARCOS satellite (Perryman 1989) has provided high quality parallaxes
for $\sim 118,000$\ stars. These data can be used to address a variety of
different astrophysical problems. HIPPARCOS parallaxes for about one hundred
nearby dwarfs with metal abundances in the range $-2.5<$[Fe/H]$<0.2$\  were
available to us, before their release to general public, as result of the
Hipparcos FAST proposal n.022. These objects had been originally selected
because they seemed the most suited for globular cluster distance and age
derivations via main sequence fitting technique. They have already been used
in a careful revision of the age of the oldest globular clusters, Gratton et
al. (1997a), who obtained ages consistent with inflationary models for the
universe, if the value of the Hubble constant is in the range 50-75
km~s$^{-1}$~Mpc$^{-1}$.

Besides accurate parallaxes, an appropriate determination of {\bf magnitudes},
{\bf colours} and {\bf chemical composition} of the dwarf sample were crucial
in the derivation of distances and ages of GCs via main sequence fitting
(Gratton et al., 1997a). The relation between colour and magnitude of main
sequence stars near the turnoff is very steep: $4<dV/d(B-V)<7$. Thus when
fitting the main sequence of GC's with sequences of local subdwarfs, an error
of only 0.01 mag in the colours translates into an uncertainty of $\sim 1$ Gyr
in the derived cluster ages, $\sim$ 50 K in  the derived effective
temperatures, and, in turn, of 0.04-0.05 dex in the derived abundances.
Moreover, a basic assumption of Gratton et al's (1997a) age derivation was
that the nearby subdwarfs share the same chemical composition of main sequence
stars in globular clusters. This assumption was verified by a homogenous
spectroscopic analysis of both cluster and field stars. Abundance
determinations for giants in globular cluster were published by Carretta \&
Gratton (1997); cluster main sequence stars are too faint for a reliable
analysis until more 8~m class telescopes become available. Here, we present the
parallel spectroscopic study for 66 of the 99 nearby dwarfs in our sample.

The main purposes of this study were:
\begin{itemize}
\item to provide a photometric data-base for the programme stars to be used
both in the Globular Cluster distance and age derivation and in the abundance
analysis. An accurate and homogeneous photometric data-base for these stars is
needed for a number of reasons: first because the determination of reliable
absolute magnitudes directly bears upon the availability of accurate apparent
magnitudes; second because $B-V$, and possibly $V-I$ colours (the very deep,
high resolution, colour magnitude diagrams recently obtained for a number of
GCs by HST, are in a photometric band similar to the I band) are required to
build up the subdwarf template sequences to compare with the Globular Clusters
main sequences; and third because colours ($B-V$, $V-K$, $b-y$ and possibly
$V-I$) are necessary to derive the effective temperatures used in abundance
analyses. The vast data available in the literature should be homogenized,
compared and implemented with the photometric data collected from the
Hipparcos/Tycho mission, and transformed to a common uniform photometric
system.
\item to provide metal abundances for the programme stars. The analysis should
be homogenous with the globular cluster one, to avoid spurious errors in the
derived distances and ages; accurate because one anticipates that only a few
stars will finally be used in the main sequence fitting (once metal-rich,
binary, and too-distant stars are discarded); and should use updated model
atmospheres (Kurucz 1993), in order to have consistent results for both the
Sun and the programme stars (else systematic errors will be introduced in the
globular cluster ages when turnoff magnitudes are compared to isochrones)
\item to derive abundances not only for Fe, but also for O and for the
$\alpha-$elements (mainly Mg and Si; but also Ca and Ti), because derived ages
are affected at a significant level by non-solar abundance ratios. Many
studies beginning with the pioneering study by Wallerstein et al. (1963),
through the early echelle surveys of Cohen, Pilachowski, and Peterson (see
Pilachowski, Sneden \& Wallerstein, 1983, and references therein) to recent
large-sample abundance determinations in individual clusters (reviewed by e.g.
Kraft 1994; Carretta, Gratton \& Sneden, 1997) have shown that these elements
are overabundant in globular cluster stars. Available data indicate that a
similar overabundance is shared by metal-poor field stars (see e.g. Wheeler,
Sneden \& Truran 1989). However, it is important to verify that this is the
case also for the subdwarfs used in the main sequence fitting procedure, since
a few subdwarfs are known to have no excess of heavy elements (King 1997,
Carney et al. 1997).
\item to determine high precision ($<1$~km/s error) radial velocities for the
target stars. These velocities can be compared with values in the
literature, thus providing useful information on possible unknown binaries
present in our sample. Binary contamination is one of the major
concerns in the derivation of distances to globular clusters using the
subdwarf main sequence fitting method.
\item to compare stellar gravities deduced from Hipparcos parallaxes with
those from ionization equilibrium computations. Several authors have suggested
that there may be an appreciable Fe overionization in the atmosphere of late
F-early K-dwarfs (Bikmaev et al., 1990; Magain \& Zhao, 1996; Feltzing \&
Gustafsson, 1998). If this were true, Fe abundances derived assuming LTE would
be largely in error. Gratton et al. (1997b) have shown that it is difficult to
reconcile a large Fe overionization in cool dwarfs with the rather good
ionization equilibrium found for lower gravity, warmer stars (see e.g. the
case of RR Lyrae stars: Clementini et al. 1995). But now, thanks to Hipparcos,
we are able to derive accurate surface gravities directly from luminosities
and masses (the last ones using stellar evolution models) for a large sample
of metal-poor dwarfs; we can then test whether  a significant Fe
overionization actually exists in these stars by simply comparing the
abundances provided by neutral and singly ionized lines.
\item Finally, to use this large, homogenous data set to recalibrate
photometric and low $S/N$\ spectroscopic abundances, that may be used to obtain
moderate accuracy (errors $\sim 0.15$~dex) abundances for thousands of
metal-poor stars. In particular we have tied to our scale the metal abundances
by Schuster \& Nissen (1989), Carney et al. (1994), and Ryan \& Norris (1991).
\end{itemize}

\section{Basic data for the Subdwarfs}

\subsection{Photometric Data}

Photometric data ($UBVRIJHK$ and Str\"omgren $b-y$, $m_1$\ and $c_1$) for all
the programme stars (99 objects) have been collected from a careful selection
of the literature data available in the SIMBAD data-base. A few stars (five
objects) were re-observed. The Hipparcos catalog provides $V$ magnitudes and
$B-V$, $V-I$ colours (in the Johnson-Cousins system) for all our objects. All
$V-I$ colours are from the literature, while $V$ and $B-V$ are either from the
literature or measured from the Hipparcos/Tycho missions (Grossmann et al.,
1995). In particular, 55 of our objects have $V$\ magnitudes directly measured
by Hipparcos ($V_H$), and 24 stars have $B-V$ colours measured by Tycho,
$(B-V)_T$.

Data from different sources and in different photometric systems were
transformed to a uniform photometric system, using equations which are
available in electronic form upon request to the first author.  Average
magnitudes and colours for the programme stars were derived combining
literature and Hipparcos data according to an accurate standardization
procedure that  is briefly outlined below. Final adopted magnitudes and
colours for the programme stars are given in Table 1 where $V$, $B-V$, and
$V-K$ are in the Johnson system, and the $V-I$ and $b-y$ colours are in
Cousins and Str\"omgren photometric systems, respectively. Uncertain values
are marked by a colon. Only the first 20 entries of the table are given in the
paper; data for the remaining stars are available in electronic form.

\subsubsection{$V$\ magnitudes and $B-V$ colours}

Mean $V$\ magnitudes ($V_{\rm g}$) and $B-V$ colours, $(B-V)_{\rm g}$, were
computed as the average of the independent literature data. These mainly
consist of Johnson $UBV$ photometry by Sandage \& Kowal (1986), Carney (1978,
1983a, 1983b), Carney \& Latham (1987), Laird, Carney \& Latham (1988), and
Carney et al. (1994). According to the distribution of individual measures
around the average, measures that were more than 0.07 and 0.03 mag off the
mean V and $B-V$ values, respectively,  were discarded. The ($V_{\rm g}$) and
$(B-V)_{\rm g}$ values thus obtained were compared with $V_{\rm H}$'s for the
55 objects with Hipparcos $V$ magnitudes and $(B-V)_{\rm T}$'s for the 24
objects with Tycho colours, respectively. The best fit regression lines are:
$V_{\rm H} = 1.003 V_{\rm g} - 0.038$ and $(B-V)_{\rm T} = 1.002(B-V)_{\rm g}
+ 0.012$.

$V_{\rm H}$'s  were transformed to $V_{\rm g}$'s using the regression equation
given above, and then averaged with the ground-based measures; the mean values
so derived are listed in column 7 of Table 1. The average uncertainty of these
mean values is 0.014~mag (average on 87 objects).

\begin{figure}
\vspace{5.5cm}
\includegraphics{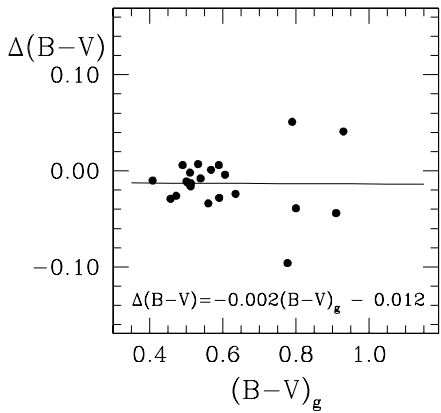}
\caption{ Residuals $\Delta (B-V)=(B-V)_{\rm g} - (B-V)_{\rm T}$ of the mean
$B-V$ colours estimated from the literature data, $(B-V)_g$, and Tycho'
$B-V$'s, $(B-V)_{\rm T}$, for the 24 objects with Tycho measures, plotted
versus $(B-V)_{\rm g}$} 
\label{f:figure01} 
\end{figure}

Figure~\ref{f:figure01} shows the residuals $(B-V)_{\rm g} - (B-V)_{\rm T}$ versus
$(B-V)_{\rm g}$. A zero point offset of $-$0.01~mag is present in this plot
and a strong scatter for  colours redder than 0.75~mag. However, we only have
5 objects with $B-V>0.75$ in this plot, so it is difficult to assess the
actual meaning and relevance of this scatter. We have thus resolved to use
Tycho's $B-V$'s only if they did not differ more than the residual of the mean
and, in any case, no more than 0.015~mag. This is true for 13 out of the 24
objects with $(B-V)_{\rm T}$. New mean $B-V$'s were computed as the average of
the independent literature data and the $(B-V)_{\rm T}$'s for these objects.
The adopted $B-V$'s are listed in column 10 of Table 1. The average
uncertainty of these values is 0.011~mag (average on 73 objects).


\subsubsection{$V-K$ colours}

Forty-seven of our objects have $K$\ magnitudes published in the literature.
These mainly include $K$\ photometry in the CIT system (Carney 1983b, Laird
1985, Laird, Carney \& Latham (1988), in the TCS system by Alonso, Arribas \&
Martinez-Roger (1994) and in the OT system by Arribas \& Martinez Roger (1987).
There are also a few data in the Johnson, OT, KPNO, OAN, AAO and Glass (1974)
systems. $V-K$ colours in the various photometric systems were formed from the
$V$ values listed in column 7 of Table 1, transformed to Johnson and then
averaged. The mean $V-K$'s in the Johnson system so derived are listed in
column 12 of Table 1. The average uncertainty of these values is 0.016~mag
(average on 24 objects).

\subsubsection{$V-I$ colours}

\begin{figure}
\vspace{5.5cm}
\includegraphics{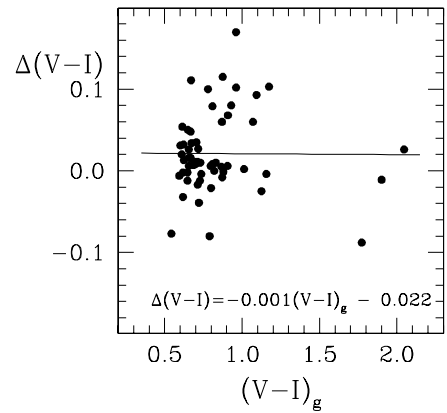}
\caption{ Residuals $\Delta (V-I)=(V-I)_g - (V-I)_H$ of the mean  $(V-I)$
colours estimated from the literature data, $(V-I)_g$, and those listed in the
Hipparcos catalog $(V-I)_H$, plotted versus $(V-I)_g$} 
\label{f:figure02} 
\end{figure}

In order to check whether we could build up subdwarf template main sequences
in $V-I$, we also compiled the available photometry in $R-I$\ and $V-I$\
colours. We found literature $R$ and $I$ photometric data for 59 of our stars.
The literature sources are Carney (1983a,b), Laird (1985), Bessel (1990),
Cutispoto (1991a,b) and unpublished, Upgren (1974), Eggen (1978) and Weis
(1996). Data are in three different photometric systems: Cousins, Johnson, and
Kron (Cousins 1976a,b, Johnson et al 1966, Kron, White \& Gascoigne 1953).
$V-I$ colours directly in the Cousins system (hereinafter $(V-I)_{\rm Cg}$)
are available only for 27 of our targets. For the remaining objects the
literature $(R-I)_{\rm K}$, $R_{\rm K}$, $(R-I)_{\rm J}$, and $R_{\rm J}$
values were combined with the $V$ values in Table 1 to form $(V-I)_{\rm K}$'s
and $(V-I)_{\rm J}$'s; these were then transformed to Cousins using Weis
(1996) and Bessell (1979) relations respectively. Finally, average $(V-I)_{\rm
g}$ in the Cousins system were derived as the mean of all the three sets of
values; they are listed in Column 13 of Table 1. The average uncertainty of
these values is 0.022~mag (average on 24 objects). The Hipparcos catalog lists
$V-I$ values in the Cousins system for all our stars. The $(V-I)_{\rm g}$ were
compared to $(V-I)_{\rm H}$ getting the regression line $(V-I)_{\rm H} =
0.996(V-I)_{\rm g} - 0.019$. The residuals, $(V-I)_{\rm g} - (V-I)_{\rm H}$,
versus $(V-I)_{\rm g}$ are shown in Figure~\ref{f:figure02}.

\begin{table*}
\begin{minipage}{160mm}
\begin{center}
\caption{Hipparcos parallaxes and colours for the programme stars}
\end{center}
\scriptsize
\begin{tabular}{rrrlrcrcccccccc}
\hline\hline
 HIP~~&LTT~~&HD~~&~Gliese&$\pi$~~~&$\delta \pi/\pi$&$V~$&$M_v$&$\sigma (M_V)$&$B-V$&$b-y$
&$V-K$&$V-I$&N&Comments  \\
        &   &  &      &(mas)&                &   &     &              &     &     
&    &   & \\
\hline
   999&10065&        &G030-52&24.69  &0.049& 8.46&5.43&0.10 &0.790:&0.498
&2.183&0.960&2&SBO(2),TP(2)\\
  1897&10137&        &G032-16&17.09  &0.080& 9.68&5.84&0.17 &0.880&0.519&     &      &0& \\
  3985&10310&    4906&G032-53&~9.04  &0.153& 8.76&3.54&0.31 &0.780&0.483&     &      &3& \\
  6012&     &    7783&       &10.02  &0.191& 9.42&4.42&0.38 &0.660&     &     &      &0& \\
  6037&     &    7808&       &33.18  &0.062& 9.76&7.36&0.13 &1.000&0.572&     &1.125 &0& \\
  6159&730  &    7983&G271-34&14.91  &0.082& 8.90&4.76&0.17 &0.585&0.387&1.515&0.705 &0& \\
  6448&10502&    8358&G071-03&15.21  &0.063& 8.28&4.19&0.13 &0.723&0.466&2.014
&0.875 &2&FR(1,2),SB2P(2)\\
  7217&10541&    9430&G034-36&15.33  &0.081& 9.04&4.96&0.17 &0.625&     &     
&      &2&SSB(2)\\
  8798&1007 &   11505&G071-40&26.56  &0.040& 7.42&4.55&0.08 &0.634&0.401&     
&0.711 &1&S?(1)\\
 10140&10733&        &G074-05&17.66  &0.073& 8.76&4.99&0.15 &0.576&0.390&1.524&0.708:&4& 
\\
 10652&10774&   14056&G004-10&14.43  &0.092& 9.05&4.85&0.19 &0.620&0.404&     &      &4& \\
 10921&10794&        &G073-44&18.40  &0.075& 9.12&5.44&0.16 &0.790&0.470&     
&      &2&SBP(2)\\
 12306&10869&   16397&G036-28&27.89  &0.040& 7.36&4.58&0.09 &0.588&0.385&1.485&0.666 &2& \\
 13366&10934&   17820&G004-44&15.38  &0.090& 8.39&4.32&0.19 &0.546&0.377&     
&0.674:&2&S?(1)\\
 13631&10956&        &G004-46&17.36  &0.084& 9.75&5.95&0.17 &0.800&     &     &      &0& \\
 14401&17462&        &G078-14&18.42  &0.077& 9.71&6.04&0.16 &0.777&0.465&     &0.879 &0& \\
 14594&11021&   19445&G037-26&25.85  &0.044& 8.06&5.12&0.09 &0.457&0.355&1.361&0.614 &3& \\
 14992&     &        &G076-68&15.89  &0.115& 9.84:&5.85:&0.24 &0.935&0.552& 2.486:&  &0& \\
 15394&11087&   20512&G005-27&17.54  &0.073& 7.42&3.64&0.15 &0.800&0.485&     
&0.830 &2&SBP(2)\\
 15797&11113&        &G078-33&39.10  &0.032& 8.96&6.92&0.07 &0.980&0.558&     
&      &2&SSB(1)\\
\hline
\end{tabular}
\medskip
\medskip

FR=Fast rotator, SSB=Suspected spectroscopic binary, VB=Visual binary,
SBP=Spectroscopic binary with preliminary orbital solution, SB2P=Double-lined
spectroscopic binary with preliminary orbital solution, SB2=Double-lined
spectroscopic binary with orbital solution, SBO= Spectroscopic binary with
orbital solution, TP=multiple system with preliminary orbital solution,
S?=suspected spectroscopic binary detected on the basis of the comparison with
Carney et al. (1994; see Section 2.4); (1) this paper, (2) Carney et al.
(1994), (3) Peterson et al. (1980).

\end{minipage}
\end{table*}


\begin{figure}
\vspace{5.5cm}
\includegraphics{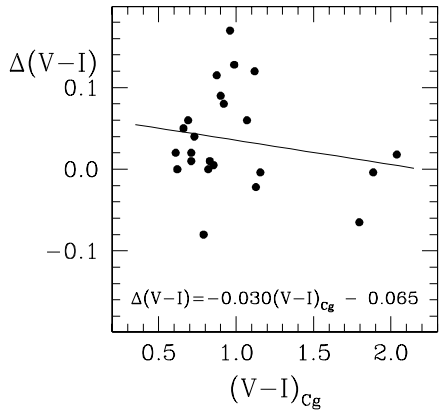}
\caption{Comparison between Hipparcos $(V-I)$'s and Cousins literature
$(V-I)$'s, $(V-I)_H$ and $(V-I)_{Cg}$, respectively, for the 27 objects
directly measured in Cousins system} 
\label{f:figure03} 
\end{figure}  


A very large scatter and a zero point offset of +0.02~mag is present in this
plot. To investigate whether at least part of this scatter might be caused by
the transformations between photometric systems, in Figure~\ref{f:figure03} we
compare the Hipparcos $(V-I)$'s and the Cousins literature $(V-I)$'s, for the
27 objects directly measured in Cousins system.

A trend and a zero point offset of $-0.065$~mag are present in this plot.
Given this overall large uncertainty we decided not to use the $V-I$ colours
in the present study as well as in Gratton et al. (1997a) study, and suggest
some caution in using the $V-I$ colours listed in the Hipparcos catalog.

\subsubsection{$b-y$ colours}

Str\"omgren $b-y$ colours for 92 of the programme stars have been published by
Laird, Carney \& Latham (1988), Carney (1983b), Laird (1985), Olsen (1983,
1984, 1994a,b), Schuster \& Nissen (1988), Schuster, Parrao \& Contreras
Martinez (1993), Anthony-Twarog \& Twarog (1987), Twarog \& Anthony-Twarog
(1995), and in the Eggen system by Eggen (1955, 1956, 1968a,b, 1972, 1978,
1979, 1987a,b). Mean $b-y$'s  were computed as the average of the independent
data available in the literature (see Column 11 of Table 1). $b-y$ values in
Eggen photometric system were also averaged since $(b-y)_{\rm E}$=$b-y$ (Eggen
1976).

\subsection{Parallaxes}

HIPPARCOS parallaxes for the programme stars are given in column 5 of Table 1.
The parallaxes as well as the absolute magnitudes $M_V$\ listed in this table
do not include Lutz-Kelker corrections (Lutz \& Kelker 1973). These
corrections depend on the distribution of the parallaxes of the population
from which the observed sample is extracted. We refer to Gratton et al.
(1997a) for a thorough discussion of this point.


\begin{figure*}
\vspace{12cm}
\includegraphics{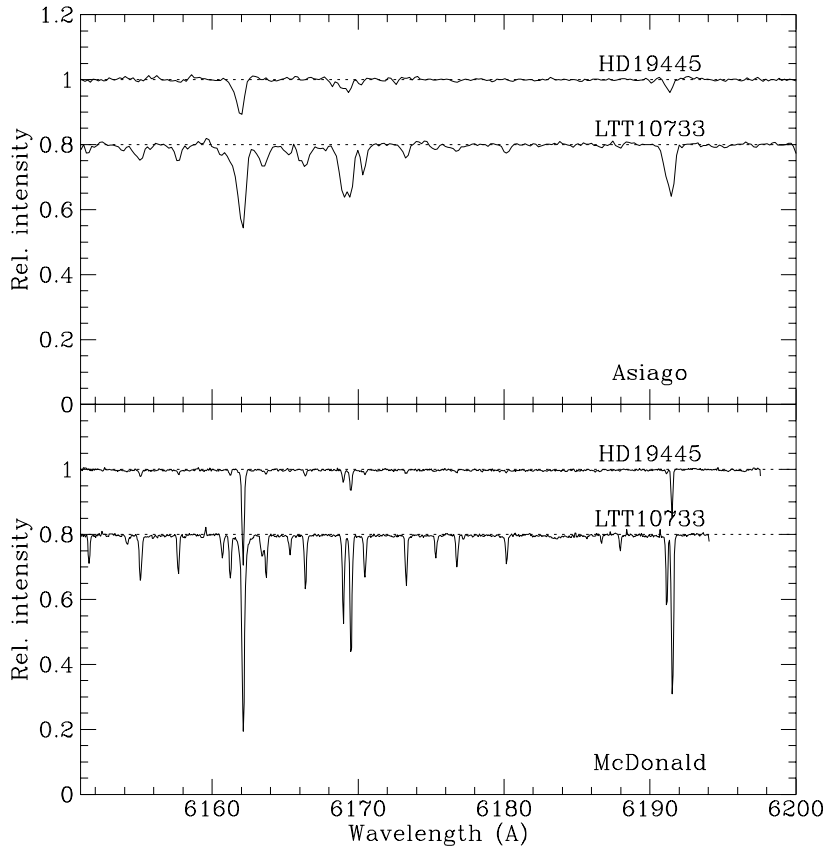}
\caption{ Portions of the spectra of HD19445 (HIP 14594) and LTT10733 (HIP 10140)
obtained with the Asiago (top panel) and McDonald (bottom panel) telescopes, 
respectively. The spectra of
LTT10733 were shifted vertically by 0.2 for a more clear display. Note the large S/N ratio of both
sets of spectra, and the
higher resolution of the McDonald spectra} 
\label{f:figure04} 
\end{figure*}  


\subsection{Spectroscopic data: Observations and Reductions}

High dispersion spectra for the programme stars were acquired using the 2d
echelle coud\`e spectrograph at the 2.7~m telescope at McDonald and the REOSC
echelle spectrograph at the 1.8~m telescope at Cima Ekar (Asiago), during the
years 1994, 1995 and 1996. McDonald spectra have a very high quality (resolution
R=60,000, $S/N\sim 200$, spectral coverage from about 4,000 to 9,000 \AA);
they are available for 22 stars (most of them with [Fe/H]$<-0.8$). The Cima
Ekar telescope provided somewhat lower quality spectra (resolution R=15,000,
$S/N\sim 200$, two spectral ranges 4,500$< \lambda <$7,000 and  5,500
$<\lambda<$8,000~\AA) for 58 stars. There are 14 stars in common between the
two samples. Portions of the spectra of some of our targets taken with the 2
different instrumental configurations are shown in Figure~\ref{f:figure04}.
Spectral ranges were selected in order to cover a large variety of lines,
including the permitted OI triplet at 7771-7774~\AA, which is the only feature
due to O easily measured in the spectrum of the metal-poor dwarfs. We collected
spectra for 66 of the objects listed in Table 1. The number of spectra
available for each object is given in column 14 of the Table. Some of the
spectra were not useful to measure abundances. For example, HIP 46191 turned
out to be a double-lined spectroscopic binary, while HIP 999, 6448 and 116005
spectra have very broad lines suggesting that they may either be fast rotating
objects or double-lined spectroscopy binaries not resolved in our
spectroscopy. Carney et al. (1994), published data for HIP 999 and 6448. They
flagged HIP 999 as a spectroscopic binary with an orbital solution and as a
multiple system with a preliminary orbital solution. Only HIP 6448, which they
classify as a double-lined spectroscopic binary with preliminary orbital
solution, is included among the "rapid rotating objects" (see their Table 9).
Those stars, as well as other known or suspected binaries present in our
sample have been flagged in Column 15 of Table 1. Finally, 4 of the stars in
Table~1 (namely HIP 17147, 38625, 76976 and 100568) have abundances from high
resolution spectroscopy recently published by Gratton, Carretta \& Castelli
(1997); these objects were not reobserved.
 
Bidimensional spectra provided by the large format CCD detectors were reduced
to unidimensional ones using standard routines in the IRAF package. Next steps
were performed using the ISA package written by one of the authors (R.G.G.) and
running on PCs. Considerable care was devoted to the somewhat subjective
reduction to a fiducial continuum level. The bluest orders
($\lambda<4,900$~\AA) were not used; however, assignment of a fiducial
continuum is still difficult on the Asiago spectra of the coolest and most
metal rich stars. To reduce this concern, in the final analysis we rejected all
lines for which the average value of the spectrum $c$\ (normalized to the
fiducial continuum) is smaller than 0.9 over a region having a 200 pixels
width ($\sim \lambda/200$) centered on the line.

Equivalent widths $EW$s of the lines were measured by means of a Gaussian
fitting routine applied to the core of the lines; they are available in
electronic form from R. Gratton. This procedure neglects the contribution of
the damping wings, which are well developed in strong lines in the spectra of
dwarfs (given the higher resolution, the effect is more evident in the McDonald
spectra than in the Asiago ones). However, fitting by Voigt profiles would
make the results very sensitive to the presence of nearby lines and to even
small errors in the location of the fiducial continuum, a well known problem
in solar analysis (see e.g. the discussion in Anstee, O'Mara, \& Ross 1997). A
full analysis would have required a very time consuming line-by-line
comparison with synthetic spectra. We deemed it beyond the purposes of the
present study, although the high quality of the McDonald spectra makes this
effort worthwhile in future studies. Here, we simply applied an average
empirical correction to the $EW$s of strong lines ($EW>80$~m\AA). This
correction was obtained from a comparison of $EW$s measured using the Gaussian
routine and by direct integration for both the clean Ca~I line at 6439~\AA\ and
synthetic spectra of typical Fe lines. Since the corrections depend on the
instrumental profile (and hence on the resolution), individual corrections
were derived for the Asiago and McDonald spectra. In the end, to avoid use of
large (and hence uncertain) corrections, only lines with $\log
{EW/\lambda}<-4.7$\ were used in the final analysis (corrections to the $EW$s
for these lines are $\leq 7$~m\AA, which is well below 10\%). We also dropped
weak lines ($\log {EW/\lambda}<-5.7$) measured on Asiago spectra, since they
were deemed too close to the noise level.

The large overlap between Asiago and McDonald samples (13 stars in common,
after the short period double-lined spectroscopic binary HIP 48215 is eliminated), allowed
standardization of the equivalent widths used in the analysis. Our procedure
was to empirically correct the Asiago $EW$s to the McDonald ones. The final
correction (based on 346 lines) is: 
$$EW_{\rm final} = (1.079\pm 0.020)~EW_{\rm original}+(42\pm 28)(1-c)-5.5~~
{\rm m\AA}$$ 
The r.m.s. scatter around this relation is 7.8~m\AA.

External checks on our $EW$s are possible with Edvardsson et al. (1993) and
Tomkin et al. (1992). Comparisons performed using McDonald $EW$s alone show
that they have errors of $\pm 4$~m\AA. From the r.m.s. scatter between Asiago
and McDonald $EW$s, we then estimate that the former have errors of $\pm
6.7$~m\AA. When Asiago and McDonald $EW$s are considered together, we find
average residuals (this paper$-$others) of $-0.2\pm 1.0$~m\AA\ (39 lines, r.m.s.
scatter 6.1~m\AA) and $+0.8\pm 1.0$~m\AA\ (36 lines, r.m.s. scatter 5.9~m\AA)
with Edvardsson et al. (1993) and Tomkin et al. (1992), respectively. 
Our $EW$s are on the same system of these two
papers. 

\subsection{Radial Velocities: Searching for Unknown Binaries}

Radial velocities (RV's) were measured from the spectra of our stars (62
objects, once the 3 objects with very broad lines HIP 999, 6448 and 116005,
and the double-lined spectroscopic binary HIP 46191 are discarded).
Forty-seven of these objects have multiple observations, but most are
consecutive exposures and thus cannot give useful information about unknown
binary systems contaminating the sample. Average RV's (with individual values
weighted according to their $\sigma$), have been derived for the objects with
multiple observations. Radial velocities are natural by-products of the EWs
measurements, since the Gaussian fitting routine used to measure the EWs also
measures the radial velocity of the centroid of the lines. About 50 - 100
lines were measured in each star and the zero point of our radial velocities
was set by measuring $\sim 10$\ telluric lines present in the spectra.

\begin{table}
\scriptsize
\caption{Radial Velocities}
\begin{tabular}{rrrrrc}
\hline\hline
\multicolumn{1}{r}{HIP}&
\multicolumn{1}{c}{HD/Gliese}&
\multicolumn{1}{c}{~RV}&
\multicolumn{1}{c}{$\sigma$}&
\multicolumn{1}{c}{$\Delta$RV}&Comments\\
&\multicolumn{1}{c}{(this paper)}&
\multicolumn{1}{c}{(this paper)}&
&\\
\multicolumn{1}{c}{}&\multicolumn{1}{c}{}&\multicolumn{1}{c}{(km/s)}&\multicolumn{1}{c}{(km/s)}&
\multicolumn{1}{c}{(km/s)}&\\
\hline
3985 &   4906&   $-$83.2~~~~~& 1.6~~~~~~ &   0.0~~~~&\\
7217 &   9430&   $-$54.7~~~~~& 0.2~~~~~~ &$-$1.3~~~~&SSB\\
8798 &  11505&   $-$13.8~~~~~&           &   3.0~~~~&*\\
10140&G074-05&      27.4~~~~~& 1.2~~~~~~ &   0.3~~~~&\\
10652&  14056&   $-$21.8~~~~~& 0.4~~~~~~ &$-$0.7~~~~&\\
10921&G073-44&      43.3~~~~~& 0.0~~~~~~ &$-$1.2~~~~&SBP\\
12306&  16397&  $-$100.0~~~~~& 0.5~~~~~~ &$-$0.1~~~~&\\
13366&  17820&       5.0~~~~~& 0.2~~~~~~ &$-$1.3~~~~&*\\
14594&  19445&  $-$140.0~~~~~& 0.3~~~~~~ &   0.5~~~~&\\
15394&  20512&     7.2~~~~~& 0.4~~~~~~ &$-$1.2~~~~&SBP\\
15797&G078-33&       7.2~~~~~& 2.9~~~~~~ &          &*\\
16169&  21543&     64.4~~~~~& 0.6~~~~~~ &   0.9~~~~&\\
16788&  22309&   $-$28.2~~~~~& 0.6~~~~~~ &$-$0.7~~~~&\\
20094&  27126& $-$44.8~~~~~& 0.3~~~~~~ &$-$1.9~~~~&*\\
21272&  28946& $-$46.2~~~~~& 1.3~~~~~~ &          &\\
21703&  29528& $-$19.3~~~~~& 0.6~~~~~~ &$-$0.3~~~~&\\
29759&G098-58&     242.2~~~~~& 1.4~~~~~~ &0.0~~~~&SSB\\
30862&  45391&    $-$6.1~~~~~& 0.2~~~~~~ &          &\\
34414&  53927&     18.9~~~~~& 0.2~~~~~~ &          &\\
35377&  56513& $-$34.2~~~~~& 0.5~~~~~~ &          &\\
36491&  59374&      90.5~~~~~& 0.2~~~~~~ &$-$0.4~~~~&\\
37335&G112-36&      49.6~~~~~& 0.6~~~~~~ &   0.4~~~~&\\
38541&  64090&  $-$234.8~~~~~& 0.1~~~~~~ &   0.2~~~~&SSB\\
48215&  85091&      32.7~~~~~&28.5~~~~~~ &$-$10.8~~~~&SBO\\
49615&  87838&      23.1~~~~~& 1.2~~~~~~ &$-$0.1~~~~&\\
49988&  88446&      61.0~~~~~& 0.3~~~~~~ &$-$0.6~~~~&\\
50139&  88725& $-$22.1~~~~~& 0.9~~~~~~ &$-$0.1~~~~&\\
53070&  94028&     66.1~~~~~& 0.4~~~~~~ &   0.7~~~~&SSB\\
54196&  96094&     0.5~~~~~& 0.1~~~~~~ &$-$0.1~~~~&\\
57939& 103095& $-$98.9~~~~~&           &$-$0.5~~~~&SSB\\
60956& 108754&     0.4~~~~~& 1.6~~~~~~ &$-$0.3~~~~&SBO\\
62607& 111515&       3.9~~~~~& 1.3~~~~~~ &   1.5~~~~&SSB\\
64115& 114095&     77.8~~~~~&           &          &\\
64345& 114606&     26.0~~~~~& 0.3~~~~~~ &$-$0.7~~~~&\\
64426& 114762&     50.5~~~~~& 0.5~~~~~~ &   1.2~~~~&SBO\\
65982& 117635&  $-$50.8~~~~~& 0.1~~~~~~ &   0.7~~~~&SSB\\
66509& 118659&  $-$44.7~~~~~& 1.5~~~~~~ &   0.6~~~~&\\
66860& 119288&  $-$10.9~~~~~& 0.3~~~~~~ &          &\\
72998& 131653&  $-$67.8~~~~~&           &          &\\
74033& 134113&  $-$58.7~~~~~& 0.1~~~~~~ &   1.9~~~~&SBP\\
74234& 134440&    311.1~~~~~&           &$-$0.4~~~~&\\
74235& 134439&    310.5~~~~~&           &$-$0.1~~~~&\\
80837& 148816&  $-$47.6~~~~~& 0.5~~~~~~ &   0.3~~~~&\\
81170& 149414& $-$152.1~~~~~&           &  17.7~~~~&SBP\\
81461& 149996&  $-$36.6~~~~~&           &$-$0.7~~~~&\\
85007& 157466&     34.8~~~~~& 0.4~~~~~~ &          &\\
85378& 158226&  $-$73.5~~~~~&           &$-$0.1~~~~&\\
85757& 158809&      4.5~~~~~&           &   0.7~~~~&SSB\\
92532& 174912&  $-$13.0~~~~~& 1.1~~~~~~ &   0.2~~~~&\\
95727& 231510&      5.5~~~~~& 0.4~~~~~~ &   0.7~~~~&\\
96077& 184448&  $-$21.7~~~~~&           &   0.4~~~~&\\
97023& 186379&   $-$8.6~~~~~& 0.1~~~~~~ &          &\\
97527& 187637&   $-$0.4~~~~~&           &          &\\
100792&194598& $-$248.2~~~~~& 0.4~~~~~~ &$-$0.5~~~~&\\
103987&200580&      4.6~~~~~& 0.6~~~~~~ &   6.1~~~~&TP\\
104659&201891&  $-$44.7~~~~~& 1.3~~~~~~ &$-$0.2~~~~&\\
105888&204155&  $-$84.5~~~~~&           &   0.1~~~~&\\
109450&210483&  $-$73.2~~~~~& 0.8~~~~~~ &          &\\
109563&210631&  $-$12.5~~~~~&           &   0.0~~~~&SBP\\
112229&215257&  $-$33.5~~~~~& 1.0~~~~~~ &   0.1~~~~&\\
112811&216179&   $-$4.3~~~~~&           &$-$0.2~~~~&\\
117918&224087&  $-$28.3~~~~~& 0.5~~~~~~ &$-$0.1~~~~&SBO\\
\hline
\end{tabular}
\label{t:velrad}
\end{table}

Carney et al. (1994) have published radial velocities for 50 of these objects;
among them 18 are confirmed or suspected binaries. A comparison of the
remaining objects in common, (28 and 13 stars for the Asiago and McDonald
samples, respectively) shows that there is a systematic shift and  a trend
between Asiago estimates and Carney et al's values described by the linear
regression RV$_{\rm this~paper}-RV_{\rm Carney}= -0.0146 RV -5.8970$, with
correlation coefficient $r=-$0.726. The trend and zero point with the McDonald
 spectra is much lower: RV$_{\rm this~paper}-RV_{\rm Carney}= -0.0076 RV
-0.4278$, $r=-$0.923. We think that the lower resolution of the Asiago spectra
is responsible for the rather large offset found between Asiago and Carney et
al's values. The radial velocities measured from our spectra, (corrected
accordingly to the equations given above, to tie them to Carney et al's
system) are listed in Table~\ref{t:velrad}. Also shown are the standard
deviations of the means (for stars with multiple observations, Column 4).

These standard deviations may give some information on the binary
contamination suffered by our sample. The quadratic mean of the rms deviations
corresponds to 0.81 km/s, after the previously mentioned 18 known or suspected
binaries are eliminated, (average on 33 objects). This value drops to 0.57
km/s, (average on 32 objects), if we also eliminate HIP 15797 which $\sigma$
is higher than 2.5 $\times$ the quadratic mean. We suspect that this object is
likely to be a previously unrecognized binary. We can conservatively assume
that our radial velocity measurements are accurate to 1 km/s.

Column 5 of Table~\ref{t:velrad} gives the residuals, this paper $-$ Carney et
al. (1994) for the stars in common. These residuals point to possible unknown
binaries contaminating the sample. Their quadratic mean, once all known or
suspected binaries are eliminated, corresponds to 0.62 (average on 32
objects). If we eliminate objects whose residuals are higher than 2.5 $\times$
the quadratic mean we obtain 0.24 (average on 30 stars, discarding HIP 8798,
20094), and 0.19 if we also eliminate HIP 13366 (average on 29 stars). In
summary, besides the 18 already known or suspected binaries
Table~\ref{t:velrad} includes 3 objects whose binarity is highly probable
(namely, HIP 8798, 15797, 20094) and 1 more object (HIP 13366) whose binarity
is likely. All these objects are marked by an asterisk in Column 6. Radial
velocities measured from the individual spectra of known and newly discovered
binaries are available in electronic form upon request to the first author.

In summary, we find that 26 out of 66 objects with spectroscopic observations
are confirmed or suspect binaries, i.e. $\sim 39 \%$ of the total sample. Of
them 20 are binaries already known and 6 are new possible binaries detected in
the present study. We refer to Gratton et al. (1997a) for a more thorough
discussion of the the binary contamination affecting our sample.

\section{Abundance Analysis}

\subsection{Atmospheric Parameters}

The abundance derivation followed precepts very similar to the reanalysis of
$\sim 300$\ field dwarfs and $\sim 150$\ globular cluster giants described in
Gratton, Carretta \& Castelli (1997) and Carretta \& Gratton (1997). The same
line parameters were adopted. The atmospheric parameters needed in the analysis
were derived according to the following procedure: 
\begin{enumerate} 
\item we assumed initial values of $\log g=4.5$\ and metal abundances derived
from the $uvby$\ photometry using the calibration by Schuster \& Nissen (1989) 
\item initial effective temperatures were derived from the $B-V$, $b-y$, and
$V-K$\ colours listed in Table 1 using the empirical calibration
of Gratton, Carretta \& Castelli (1997) for population I stars (assumed to be valid for
[Fe/H]=0), and the abundance dependence given by Kurucz (1993) models.
Following Carney et al. (1994), who assumed zero reddening values for all the
stars in our sample, colours were not corrected for reddening. 
\item a first iteration value of $\log g$\ was then obtained from the absolute
bolometric magnitude (derived from the apparent $V$\ magnitudes, the parallaxes
from Hipparcos, and the bolometric corrections $BC$\ given by Kurucz, 1993),
and the masses derived by interpolating in $T_{\rm eff}$\ and [A/H] within the
Bertelli et al. (1994) isochrones 
\item the two last steps were iterated until a consistent set of values was
obtained for $T_{\rm eff}$, $\log g$, and [A/H] 
\item the equivalent widths were then analyzed, providing new values for $v_{\rm
t}$\ and  [A/H] (assumed to be equal to the [Fe/H] value obtained from neutral
lines) 
\item the whole procedure was iterated until a consistent parameter set was
obtained. Note that only $BC$'s and masses are modified, so that convergence
is actually very fast.
\end{enumerate}

\begin{table*}
\caption{Atmospheric parameters and abundances for the programme stars}
\scriptsize
\begin{tabular}{rrcccrrcrrcrrr}
\hline\hline
HIP&~~HD/Gliese&&$T_{\rm eff}$&$\log g$&[Fe/H]&[O/Fe]&[Mg/Fe]&[Si/Fe]&[Ca/Fe]&
[Ti/Fe]&~~[$\alpha$/Fe]&~[Cr/Fe]&[Ni/Fe]\\
\\
&&&(K)& & &&&&&&&&\\ 
\hline
\\
\multicolumn{14}{c}{Single stars}\\
  3985&4906&A&5149&3.61&$-$0.65&       &    &   0.19&   0.42&    &   0.30&$-$0.04&0.00\\
 10140&G074-05&M&5755&4.38&$-$0.85&   0.44&0.40&   0.19&   0.22&0.24&   0.26&0.00&   0.02\\
 10140&G074-05&A&5755&4.38&$-$0.78&   0.18&    &   0.13&   0.19&    &   0.16&   0.02&$-$0.16\\
 10652&14056&A&5647&4.31&$-$0.58&   0.34&    &   0.14&   0.34&    &   0.24&$-$0.01&   0.06\\
 12306&16397&A&5810&4.28&$-$0.50&   0.47&    &   0.24&   0.17&    &   0.20&   0.08&$-$0.05\\
 14594&19445&M&6059&4.49&$-$1.91&   0.67&0.32&       &   0.36&0.45&   0.38&   0.01&\\
 14594&19445&A&6059&4.49&$-$2.03&       &    &       &   0.46&    &   0.46&$-$0.04&\\
 16169&21543&A&5673&4.37&$-$0.50&   0.33&    &   0.07&   0.23&    &   0.15&   0.01&$-$0.04\\
 16788&22309&A&5873&4.29&$-$0.25&   0.22&    &$-$0.07&   0.02&    &$-$0.03&$-$0.19&$-$0.15\\
 21272&28946&A&5288&4.55&$-$0.03&$-$0.26&    &   0.06&   0.09&    &   0.07&   0.11&0.00\\
 21703&29528&M&5331&4.35&   0.15&   0.02&    &   0.02&   0.11&0.22&   0.11&$-$0.12&   0.02\\
 21703&29528&A&5331&4.35&   0.03&       &    &   0.08&   0.04&    &   0.06&   0.01&   0.06\\
 30862&45391&A&5707&4.46&$-$0.37&   0.08&    &$-$0.05&   0.16&    &   0.05&   0.11&$-$0.08\\
 34414&53927&A&4937&4.66&$-$0.37&   0.07&    &   0.10&   0.10&    &   0.10&   0.02&   0.16\\
 35377&56513&A&5659&4.50&$-$0.38&$-$0.22&    &   0.09&   0.20&    &   0.14&   0.04&   0.04\\
 36491&59374&A&5903&4.44&$-$0.88&   0.25&    &   0.24&   0.30&    &   0.27&   0.03&$-$0.01\\
 37335&G112-36&A&5036&2.92&$-$0.82&   0.34&    &   0.23&   0.46&    &   0.34&$-$0.09&$-$0.06\\
 49615&87838&A&6078&4.30&$-$0.30&$-$0.17&    &   0.08&   0.13&    &   0.11&   0.24&$-$0.16\\
 49988&88446&A&5935&3.97&$-$0.36&   0.35&    &   0.14&$-$0.05&    &   0.04&$-$0.08&$-$0.18\\
 50139&88725&M&5695&4.39&$-$0.55&   0.16&0.31&   0.11&   0.22&0.34&   0.24&$-$0.01&$-$0.01\\
 50139&88725&A&5695&4.39&$-$0.57&   0.26&    &   0.25&   0.40&    &   0.33&   0.00&$-$0.05\\
 54196&96094&A&5879&3.97&$-$0.33&   0.23&    &   0.01&   0.03&    &   0.02&$-$0.13&$-$0.20\\
 57939&103095&M&5016&4.80&$-$1.30&       &    &       &   0.23&0.35&   0.29&$-$0.08&$-$0.06\\
 64115&114095&A&4741&2.69&$-$0.35&$-$0.35&    &$-$0.54&   0.39&    &$-$0.07&       &   0.02\\
 64345&114606&A&5611&4.28&$-$0.39&   0.11&    &   0.18&   0.21&    &   0.19&   0.08&$-$0.06\\
 66509&118659&A&5494&4.37&$-$0.55&   0.51&    &   0.13&   0.03&    &   0.08&$-$0.10&$-$0.10\\
 66860&119288&A&6566&4.19&$-$0.27&   0.15&    &   0.06&   0.06&    &   0.06&$-$0.14&$-$0.02\\
 72998&131653&M&5356&4.65&$-$0.63&   0.36&    &   0.30&   0.23&0.41&   0.31&   0.01&$-$0.06\\
 74234&134440&M&4879&4.74&$-$1.28&       &    &       &   0.11&0.18&   0.15&   0.06&$-$0.16\\
 74235&134439&M&5106&4.74&$-$1.30&       &    &   0.51&   0.15&0.22&   0.29&$-$0.06&$-$0.15\\
 80837&148816&M&5923&4.16&$-$0.64&   0.23&0.33&   0.23&   0.33&0.31&   0.30&$-$0.01&$-$0.03\\
 80837&148816&A&5923&4.16&$-$0.61&   0.39&    &   0.19&   0.30&    &   0.25&       &$-$0.18\\
 81461&149996&M&5726&4.14&$-$0.38&   0.09&0.23&   0.13&   0.29&0.23&   0.22&$-$0.02&$-$0.02\\
 85007&157466&M&6053&4.39&$-$0.32&$-$0.02&0.15&   0.00&   0.10&0.05&   0.08&$-$0.05&$-$0.08\\
 85007&157466&A&6053&4.39&$-$0.38&       &    &$-$0.01&   0.02&    &   0.00&$-$0.21&$-$0.13\\
 85378&158226&A&5803&4.18&$-$0.42&       &    &   0.44&   0.55&    &   0.50&   0.21&   0.04\\
 92532&174912&M&5954&4.49&$-$0.39&$-$0.10&0.03&   0.05&   0.14&0.24&   0.11&   0.02&$-$0.04\\
 92532&174912&A&5954&4.48&$-$0.29&       &    &   0.15&   0.12&    &   0.14&   0.04&   0.03\\
 95727&231510&M&5253&4.59&$-$0.44&   0.34&0.20&   0.14&   0.07&0.16&   0.14&   0.02&   0.00\\
 96077&184448&A&5656&4.20&$-$0.22&       &    &   0.19&   0.19&    &   0.19&$-$0.09&$-$0.15\\
 97023&186379&A&5894&3.99&$-$0.20&$-$0.03&    &   0.12&   0.10&    &   0.11&   0.13&$-$0.17\\
 97527&187637&A&6169&4.23&$-$0.10&       &    &   0.02&   0.04&    &   0.03&$-$0.01&$-$0.24\\
100792&194598&M&6032&4.33&$-$1.02&   0.35&0.25&   0.18&   0.28&0.19&   0.08&$-$0.10&$-$0.12\\
100792&195598&A&6032&4.33&$-$0.99&       &    &       &   0.37&    &   0.37&   0.04&$-$0.43\\
104659&201891&M&5957&4.31&$-$0.97&   0.35&0.32&   0.27&   0.31&0.34&   0.31&$-$0.07&$-$0.10\\
104659&201891&A&5957&4.31&$-$0.91&   0.47&    &   0.28&   0.26&    &   0.27&$-$0.08&$-$0.13\\
105888&204155&A&5816&4.08&$-$0.56&       &    &   0.18&   0.29&    &   0.24&$-$0.02&$-$0.08\\
109450&210483&A&5847&4.20&$-$0.03&   0.05&    &   0.09&   0.00&    &   0.04&   0.02&$-$0.11\\
112229&215257&A&6008&4.23&$-$0.66&   0.14&    &   0.03&   0.24&    &   0.14&   0.12&$-$0.01\\
112811&216179&M&5443&4.46&$-$0.66&   0.45&0.34&   0.23&   0.31&0.28&   0.29&   0.02&   0.01\\
\\
\multicolumn{14}{c}{Known or suspected binaries}\\
  7217&9430&A&5689&4.40&$-$0.34&       &    &   0.30&   0.38&    &   0.34&   0.08&$-$0.09\\
  8798&11505&A&5695&4.31&$-$0.05&       &    &   0.19&   0.09&    &   0.14&0.00&   0.03\\
  10921&G073-44&A&5267&4.40&$-$0.12&       &    &   0.19&   0.06&    &   0.13&   0.49&$-$0.09\\
 13366&17820&A&5849&4.19&$-$0.59&   0.37&    &   0.43&   0.37&    &   0.40&   0.22&$-$0.01\\
 15394&20512&A&5253&3.72&   0.10&$-$0.55&    &$-$0.07&   0.12&    &   0.03&   0.21&$-$0.05\\
 15797&G078-33&A&4734&4.68&$-$0.41&       &    &   0.27&   0.05&    &   0.16&$-$0.12&$-$0.01\\
 20094&27126&A&5538&4.39&$-$0.26&$-$0.07&    &   0.15&   0.23&    &   0.19&   0.03&   0.05\\
 29759&G098-58&M&5432&3.37&$-$1.83&   0.65&0.38&       &   0.26&0.25&   0.30&$-$0.27&$-$0.32\\
 29759&G098-58&A&5432&3.37&$-$1.84&       &    &       &   0.43&    &   0.43&$-$0.18&       \\
 38541&64090&M&5475&4.62&$-$1.49&       &0.40&       &   0.27&0.28&   0.32&$-$0.05&$-$0.02\\
 38541&64090&A&5475&4.62&$-$1.59&       &    &       &   0.33&    &   0.33&   0.08&$-$0.18\\
 48215&85091&M&5698&4.15&$-$0.29&   0.10&    &   0.03&   0.12&0.48&   0.21&$-$0.11&$-$0.17\\
 48215&85091&A&5698&4.15&$-$0.50&   0.42&    &   0.43&$-$0.02&    &   0.21&   0.09&$-$0.15\\
 53070&94028&M&6049&4.31&$-$1.32&   0.43&0.48&   0.52&   0.33&0.24&   0.39&$-$0.12&$-$0.12\\
 53070&94028&A&6049&4.31&$-$1.35&   0.42&    &   0.23&   0.19&    &   0.21&$-$0.20&$-$0.06\\
 60956&108754&A&5388&4.42&$-$0.58&   0.42&    &   0.08&   0.21&    &   0.14&   0.15&$-$0.07\\
 62607&111515&A&5446&4.49&$-$0.52&   0.29&    &   0.16&   0.04&    &   0.10&       &$-$0.07\\
 64426&114762&A&5928&4.18&$-$0.66&   0.49&    &   0.12&   0.21&    &   0.16&$-$0.15&$-$0.12\\
 65982&117635&A&5197&4.10&$-$0.48&   0.43&    &$-$0.12&   0.42&    &   0.15&       &$-$0.11\\
 74033&134113&A&5776&4.11&$-$0.66&   0.29&    &$-$0.03&   0.36&    &   0.16&   0.20&$-$0.18\\
 81170&149414A&M&5185&4.50&$-$1.14&   0.45&0.53&   0.25&   0.29&0.38&   0.36&   0.04&$-$0.02\\
 85757&158809&M&5527&4.07&$-$0.53&   0.50&0.37&   0.21&   0.29&0.38&   0.31&   0.02&$-$0.04\\
103987&200580&A&5934&3.93&$-$0.43&   0.05&    &$-$0.06&   0.10&    &   0.02&$-$0.07&$-$0.27\\
109563&210631&A&5785&4.12&$-$0.37&       &    &$-$0.02&   0.30&    &   0.14&   0.20&$-$0.12\\
117918&224087&A&5164&4.42&$-$0.25&       &    &$-$0.05&   0.05&    &   0.00&$-$0.14&$-$0.05\\
\hline
\end{tabular}
\label{t:paratm}
\end{table*}

\normalsize

We list in Table~3 the 
atmospheric parameters adopted for the programme stars, 
as well as the derived abundances for Fe,  
the $\alpha-$elements (O,
Mg, Si, Ca, and Ti) and the
Fe-group elements Cr and Ni. Column 
3 of
this table indicates the source of the spectroscopic material (A=Asiago,
M=McDonald). 

\normalsize

\subsection{Error analysis}

Errors in the atmospheric parameters used in the analysis were estimated as
follows. Random errors in $T_{\rm eff}$\ can be obtained by comparing
temperatures derived from different colours. The mean quadratic error estimated
in this way (once the different weight attributed to the colours are
considered: $B-V$: weight 1; $b-y$: weight 1; $V-K$: weight 4) is $\pm 45$~K.
Systematic errors may be larger: the $T_{\rm eff}$-scale used in this paper is
discussed in detail in Gratton, Carretta \& Castelli (1997). We will assume that systematic
errors in the adopted $T_{\rm eff}$'s are $\leq 100$~K. 

Random errors in the gravities may be directly estimated from the error in the
masses (5\%\footnote{This value includes internal as well as external errors.
In fact, independently of the adopted isochrone set, for
$M_{\odot}=0.75-0.85$, an error of $\sim 4 \times 10^{9}$ in the age leads to
uncertainties of 0.02-0.03 $M_{\odot}$ in the mass, at $M_{V}=5$ and 4
respectively. Additionally, the ability of different isochrone sets in
reproducing the Sun assures that contributions of external errors possibly due
to incorrect input physics are unlikely to be larger than an additional
$2-3\%$.}), $M_V$'s (mean quadratic error is 0.18 mag), and in the $T_{\rm
eff}$'s (0.8\%), neglecting the small contribution due to $BC$'s. Expected
random errors in the gravities are 0.09~dex. Systematic errors are mainly due
to errors in the $T_{\rm eff}$\ scale and in the solar $M_V$\ value. They are
about 0.04~dex.

Random errors in microturbulent velocities can be estimated from the residuals
around the fitting relation in $T_{\rm eff}$\ and $\log g$. We obtain random
errors of 0.47 and 0.17~km~s$^{-1}$\ for Asiago and McDonald spectra,
respectively. While systematic errors may be rather large (mainly depending on
the adopted collisional damping parameters, but also on the structure of the
atmosphere), they are less important in the abundance analysis, since the
microturbulent velocity is an empirical parameter derived so that
abundances from (saturated) strong lines agree with those provided by
(unsaturated) weak lines which are insensitive to the velocity field. On the
other hand, for this same reason the very low values we obtain for the cooler
stars should be reexamined more thoroughly before any physical meaning is
attributed to them (although convection velocities are indeed expected to be
much lower in the photosphere of K-dwarfs, at least insofar mixing length
theory is adopted: Kurucz, private communication). 

Random errors in the equivalent widths and in the line parameters significantly
affect the abundances when few lines are measured for a given specie. Roughly
speaking, these errors should scale as $\sigma/\sqrt{n}$\, where $\sigma$\ is
the typical error in the abundance from a single line (0.14~dex for the Asiago
spectra, and 0.11 dex for the McDonald ones, as derived from Fe~I lines) and
$n$\ is the number of lines used in the analysis (with $14<n<79$). However,
errors may be larger if all lines for a given element are in a small spectral
range (like e.g. O, for which only lines of the IR triplet at 7771-74~\AA\
were used), since in this case errors in the $EW$s for individual lines
(mainly due to uncertainties in the correct location of the continuum level)
are not independent from each other. Furthermore, undetected blends may
contribute significantly to errors when the spectra are very crowded: this is
expected to occur mainly for the Asiago spectra of cool, metal-rich stars.
These limitations should be kept in mind in the discussion of our abundances.

\normalsize
\begin{table*}
\caption{Random errors in the abundances}
\begin{tabular}{lcccccccc}
\hline \hline
Parameter & Unit & Error& [Fe/H]&[Fe/H](II-I)&[O/Fe]&[Si/Fe]&[Ca/Fe]&[Ti/Fe]\\
\hline
$T_{\rm eff}$   &  (K) & $\pm 45$ & 0.033& 0.056& 0.022& 0.029& 0.008& 0.015\\
$\log g$        & (dex)&$\pm 0.09$& 0.009& 0.048& 0.008& 0.016& 0.012& 0.000\\
${\rm [A/H]~(A)}$& (dex)&$\pm 0.07$& 0.007& 0.013& 0.019& 0.002& 0.001& 0.007\\
${\rm [A/H]~(M)}$& (dex)&$\pm 0.04$& 0.005& 0.008& 0.012& 0.001& 0.000& 0.005\\
$v_t$~(A)       &(km/s)&$\pm 0.47$& 0.055& 0.008& 0.032& 0.041& 0.008& 0.031\\
$v_t$~(M)       &(km/s)&$\pm 0.17$& 0.020& 0.003& 0.012& 0.015& 0.003& 0.011\\
r.m.s. lines (A)&      &          & 0.024& 0.081& 0.070& 0.083& 0.085& 0.085\\
r.m.s. lines (M)&      &          & 0.019& 0.041& 0.044& 0.057& 0.039& 0.035\\
\hline
total (A)       &      &          & 0.069& 0.110& 0.083& 0.098& 0.087& 0.092\\
total (M)       &      &          & 0.044& 0.085& 0.053& 0.068& 0.042& 0.040\\
\hline 
\end{tabular}
\label{t:err}
\end{table*}

Internal errors in the model metal abundances were simply obtained by summing
up quadratically the errors due to the other sources. Systematic errors can be
due to non-solar abundance ratios. We will reexamine this point in Section 3.4.
Table~\ref{t:err} gives the sensitivity  of the Fe abundances and of the
abundance ratios computed in this paper to the various error sources
considered above. Random errors in the Fe abundances are $\sim 0.07$ and $\sim
0.04$~dex for abundances derived from Asiago and McDonald spectra,
respectively. Systematic errors are mainly due to the $T_{\rm eff}$\ scale
(which in turn depends on the adopted set of model atmosphere): they are $\sim
0.08$~dex.

\subsection{ Errors in abundance analysis due to binarity }

\begin{figure}
\vspace{11.5cm}
\includegraphics{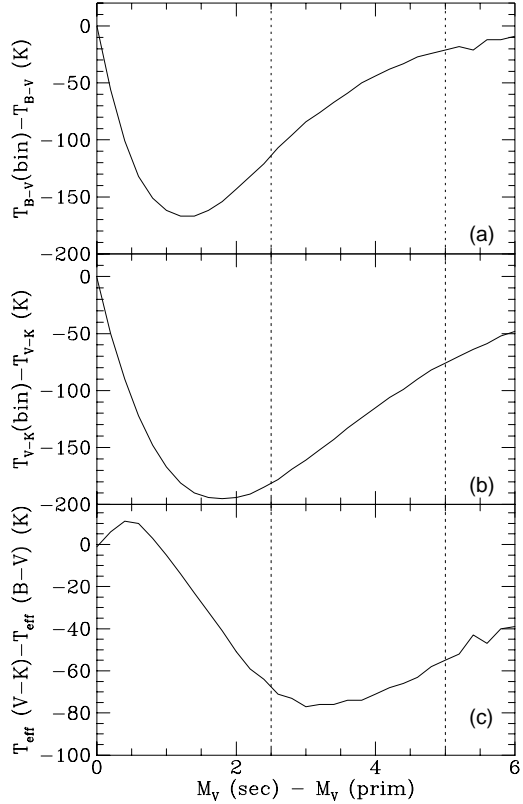}
\caption{Run of the difference between temperatures derived from $B-V$\ colours
(panel {\it a}) and $V-K$\ colours (panel {\it b}) of the combined light for a binary
system, and from the colours of the primary alone, as a function of the
magnitude difference between primary and secondary, for typical main sequence
stars (the secondary is assumed to be a faint main sequence star too).
Differences between temperatures derived from $V-K$\ and $B-V$\ are shown in
panel {\it c}. A vertical dashed line separates regions where double-lined
spectroscopic binaries are expected (magnitude differences $<2.5$~mag),
from those where binaries can be detected from their infrared excess (magnitude
differences in the range from 2.5 to 5 mag); binaries with magnitude
differences $>5$~mag are single-lined spectroscopic binaries, or visual 
binaries, and may easily go unnoticed
if extensive and accurate radial velocity observations are not available} 
\label{f:figure05} 
\end{figure}

\begin{figure}
\vspace{10.5cm}
\includegraphics{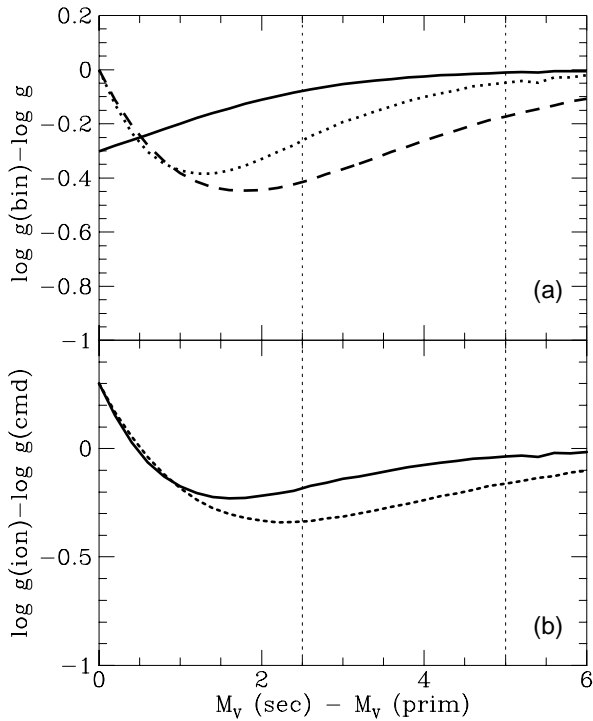}
\caption{Panel {\it a}: difference between gravities derived from the combined
light of binary systems and from the primary alone, as a function of the
magnitude difference between primary and secondary, for typical main sequence
stars (the secondary is assumed to be a faint main sequence star too). Results
obtained from the gravities derived from luminosities (via Hipparcos parallaxes)
and estimated masses (labelled log g(cmd)) are shown as a
solid line ; those obtained from ionization equilibrium of Fe are shown as
dotted (temperatures from $B-V$\ colour) and dashed lines (temperatures from
$V-K$\ colour), respectively. Panel {\it b}: run of the differences between spectroscopic
gravities and those from the location in the c-m diagram. In this case solid line
represents results for temperatures derived from $B-V$, and dotted line
results for temperatures derived from $V-K$\ colours. The vertical dashed line
has the same meaning as in Figure 5.} 
\label{f:figure06} 
\end{figure}

An additional problem in the abundance analysis is given by the presence of
known and undetected spectroscopic binaries. 
This issue is more relevant in the analysis of dwarfs than for
giants, since magnitude differences between the two components are generally
smaller for dwarfs. A large variety of possible combinations of components
exists. In the following discussion, we will only consider the case of a main
sequence primary with a smaller mass (low main sequence) secondary. This is
likely to be the most frequent combination. If neglected, binarity may affect
the abundance analysis of the primary components of such systems in various
ways: 
\begin{itemize}
\item effective temperatures, derived from the combined light, are
underestimated. Panels {\it a} and {\it b} of Figure~\ref{f:figure05} show the
run of the differences between temperatures derived from colours of the
combined light, and from colours of the primary alone, as a function of the
difference in magnitude between primary and secondary components, for typical
main sequence stars. Temperatures derived from the combined light may be as 
much as 200~K
lower than the temperature of the primary. The effect is more relevant for
$V-K$\ colours. In fact, when the magnitude difference is in the range 2-5~mag,
temperatures derived from $V-K$\ colours are lower than those derived from
$B-V$\ by more than 50~K (see panel {\it c} of Figure~\ref{f:figure05}),
raising the possibility of detecting the presence of a companion from the
infrared excess (see Gratton et al., 1997a). When the magnitude difference is 
lower than 2.5
mag there is no detectable infrared excess, but in some case the companion may
bright enough to be directly detected in the spectrum\footnote{Indeed, we
detected faint lines due to secondaries in the spectra of HIP 48215 and HIP
81170. In both cases the difference in magnitude between primary and secondary
component could be estimated as about 2.5 mag}, when phase or orbit
inclination are favourable (if separation is large there is a good chance that
the star is a known visual binary). On the other hand, when the secondary is
more than 5~mag fainter than the primary, temperatures derived from colours are
underestimated by $<20$\ and $<70$~K from $B-V$\ and $V-K$\ colours,
respectively 
\item gravities derived from luminosities (via Hipparcos parallaxes) 
and estimated masses are underestimated because the luminosity is 
overestimated (a
concurrent, smaller effect is due to the lower temperatures). Since gravities
are proportional to the inverses of luminosities, the effect may be as large as
0.3~dex, but is $<0.04$~dex when the magnitude difference between the two
components is $>2.5$~mag. On average, we expect that gravities for binaries
with magnitude differences $<6$~mag  are underestimated by $0.05-0.10$~dex,
depending on the assumed distribution of the luminosity differences between
primaries and secondaries. On the other hand, gravities derived from the
equilibrium of ionization will also be underestimated, because temperatures are
underestimated: gravities from ionization equilibrium are 0.23~dex too low if
temperatures are underestimated by 100~K. Panel {\it a} of 
Figure~\ref{f:figure06}
shows the difference between the gravity derived from the combined light and
that derived from the primary alone, as a function of the magnitude difference
between primary and secondary component, for typical main sequence stars
(again, the secondary is assumed to be a faint main sequence star). Results
obtained from the location in the colour magnitude diagrams are shown as a
solid line; those obtained from ionization equilibrium of Fe are shown as
dotted (temperatures from $B-V$\ colour) and dashed lines (temperatures from
$V-K$\ colour), respectively. Panel {\it b} of Figure~\ref{f:figure06} displays the
run of the differences between spectroscopic gravities and those from the
location in the c-m diagram. In this plot the solid line represents results for
temperatures derived from $B-V$, and the dotted line results for temperatures
derived from $V-K$\ colours. It is clear that gravities derived from ionization
equilibrium will be in most cases lower than those obtained from the c-m
diagram if a secondary component is present; the difference is larger when
$V-K$\ colours are used. On average, we expect that for binaries gravities
derived from the equilibrium of ionization should be lower than those derived
from the colour-magnitude diagram by 0.07~dex if temperatures are derived from
$B-V$, and by 0.18~dex if temperatures are derived from $V-K$. These estimates
were obtained considering only binary systems where the magnitude difference
between primary and secondary is $<6$~mag, and by assuming that secondary 
components 
distribute in luminosities as field stars (i.e. there is no correlation between
the masses of both components: see Kroupa, Tout \& Gilmore 1993) following the luminosity
function of Kroupa, Tout \& Gilmore (1993). The assumption about the luminosity
distribution of the  secondaries is not critical though: in fact, if we assume
a flat luminosity function we obtain average offsets of 0.09 and 0.20~dex in
the gravities derived using temperatures from $B-V$\ and $V-K$\, respectively. 
\item equivalent widths of lines are also affected, but a quantitative analysis
is extremely difficult, since the effect depends on the magnitude difference
between primary and secondary, on the velocity difference between the
components (in relation to the adopted spectral resolution), and on the
selected spectral region. 
When the lines from the two components are not resolved,
the equivalent widths of the primary will be slightly overestimated (lines are
stronger in the spectra of the cooler secondary), partially balancing 
the effects of the too low estimated temperatures. 
When lines from the two components are resolved,
the equivalent widths of the primary will be underestimated. However, these
general predictions may be wrong if errors in the positioning of the continuum
are also taken into account, due to the higher line density, the continuum
will generally be underestimated, thus reducing equivalent widths. 
On the whole,
errors in abundances as large as 0.2~dex may well be present when the magnitude
difference is smaller than a couple of magnitudes (errors are likely much
smaller in element-to-element abundance ratios) 
\end{itemize}

Summarizing, abundances derived for binaries are less reliable than for single
stars. We expect that our analysis will underestimate temperatures, gravities
and metal abundances for binaries. In extreme cases errors may be as large as
200~K, 0.3~dex, and 0.2~dex respectively, although typical values should be
much smaller (roughly one third of these values, on average).
In general, errors are larger when temperatures from $V-K$\ are used rather than
those from blue colours. 
We caution that corrections for individual cases
may be very different from the average ones (and, moreover, other combinations
of evolutionary stages exist). Rather than applying uncertain corrections, we
prefer to keep results obtained from known or
suspected binaries clearly distinct from those obtained from {\it bona fide}
single stars (see Table 3 and Figure 7). Of course, some {\it bona
fide} single star may indeed be a binary, if orbital circumstances were not
favourable to its detection. However, in most cases the contamination by 
possible secondary components 
is likely to be small for our {\it bona fide} single stars, and we expect that
systematic errors due to undetected binaries are on average much smaller 
than uncertainties in the temperature scale. 

\subsection{Fe abundances}

Since gravities are derived from masses and luminosities rather than from the
ionization equilibrium for Fe, we may test if the predictions based on LTE are
satisfied for the programme stars. This is a crucial test, since several
authors (Bikmaev et al. 1990, Magain \& Zhao 1996, Feltzing \& Gustafsson
1998) have suggested that Fe abundances are significantly affected by
departures from LTE in late F-K dwarfs and subdwarfs.

A proper model atmosphere analysis of Fe ionization equilibrium must take into
account the well known overabundance of O and $\alpha-$elements in metal-poor
stars (see e.g. Wheeler, Sneden \& Truran  1989, and the following subsections
of the present paper). This affects abundance derivations mainly in two ways:
(i) due to the excess of Mg and Si (which are nearly as abundant as Fe and
have similar ionization potentials) more free electrons are available
(increasing continuum opacity due to $H^-$\ and affecting Saha equilibrium of
ionization); and (ii) a stronger blanketing effect occurs (cooling the outer
layers of the atmospheres). A full consideration of these effects would
require the computation of new model-atmospheres with appropriate non-solar
abundance ratios; this is beyond the purposes of the present paper. Here to
estimate the impact of the non-solar abundance ratios we simply assumed that
the model atmosphere most appropriate for each star had metallicity scaling
down as [(Mg+Si+Fe)/H]. In practice, due to the small number of Mg and Si
lines available, we assumed [Mg/Fe]=0.38 and [Si/Fe]=0.32 for [Fe/H]$<-0.5$,
and [Mg/Fe]=$-$0.76~[Fe/H] and [Si/Fe]=$-$0.64~[Fe/H] for [Fe/H]$>-0.5$\ (as
we will see below, these are representative average values for the programme
stars). The net result of the application of these corrections for the
metal-poor stars is to increase Fe~I abundances by $\sim 0.02$~dex, and Fe~II
abundances by $\sim 0.07$~dex, the differences between Fe~I and Fe~II
abundances are then reduced by $\sim 0.05$~dex.

\begin{figure*}
\vspace{5.5cm}
\includegraphics{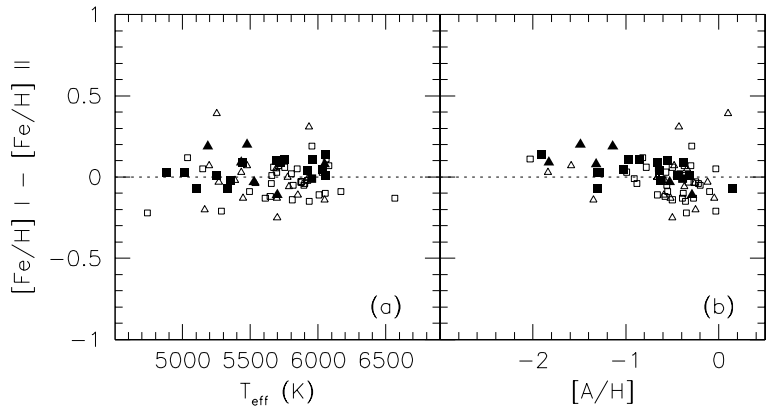}
\caption{ Run of the difference between the abundances derived from neutral and
singly ionized Fe lines, including correction for non-solar [$\alpha$/Fe] values
(see text), as a function of temperature (panel {\it a}) and overall metal
abundance (panel {\it b}). Open symbols are abundances obtained from the Asiago
spectra; filled symbols are abundances obtained from the McDonald spectra.
Squares are {\it bona fide} single stars; triangles are known or suspected
binaries } 
\label{f:figure07} 
\end{figure*}

Figure~\ref{f:figure07} displays the run of the differences between abundances
from neutral and singly ionized Fe lines, corrected for the effect of the
excess of $\alpha$-elements, as a function of effective temperature and metal
abundance.

Different symbols are used for results obtained from McDonald and Asiago
spectra and for {\it bona fide} single stars and known or suspected binaries. 
Once appropriate weights are attributed to the individual data-points of
Figure~\ref{f:figure07}, (McDonald spectra have higher weight because the
better resolution allowed us to measure a large number of Fe~II lines: $10\sim
20$, and errors in the $EW$s are smaller, while, conversely, very few Fe~II
lines could be measured in the crowded spectra of cool and/or metal-rich stars
observed with the Asiago telescope), the average difference between abundances
from Fe I and Fe II lines is only marginally different from zero. Mean
differences (FeI$-$FeII) are given in Table~\ref{t:fe1fe2m}. As expected,
results for single stars have smaller scatter than those for binaries. Also,
the smaller scatter of the higher quality McDonald spectra is evident. The
r.m.s. scatter we get for the {\it bona fide} single stars (0.07 dex for the
McDonald spectra, and 0.09 dex for the Asiago ones) agrees well with the
expected random errors in temperatures and equivalent widths (see
Table~\ref{t:err}). The lower average differences for the Asiago spectra
likely reflect some residual contamination of the few Fe~II lines measurable
on these lower resolution spectra. Reversing the results of Table~5,
we conclude that the Fe equilibrium of ionization would provide gravities on
average $0.09\pm 0.04$~dex larger than those given by masses and luminosities,
with an r.m.s. scatter of 0.13~dex for individual stars (here we only consider
results from {\it bona fide} single stars with McDonald spectra; however
results obtained from the other samples are not very different). This small
difference could be explained without invoking departures from LTE if the
adopted $T_{\rm eff}$\ scale were too high by $\sim 40$~K, well within the
quoted error bar of $\pm 100$~K.

\normalsize
\begin{table}
\caption{Mean differences between abundances given by Fe~I and II lines. For
each group of spectra the number of stars used, the average value, and r.m.s.
scatter of individual values are given }
\scriptsize
\begin{tabular}{lcccccc}
\hline \hline
Group &
\multicolumn{3}{c}{McDonald} &
\multicolumn{3}{c}{Asiago}   \\
\hline 
Single stars & 15 & $+0.04\pm 0.02$ & 0.07 & 31 & $-0.02\pm 0.02$ & 0.09 \\
Binaries     & ~6 & $+0.07\pm 0.03$ & 0.10 & 18 & $-0.01\pm 0.04$ & 0.16 \\
All          & 21 & $+0.05\pm 0.02$ & 0.08 & 49 & $-0.02\pm 0.02$ & 0.12 \\
\hline 
\end{tabular}
\label{t:fe1fe2m}
\end{table}

We therefore conclude that {\bf in our analysis the Saha ionization
equilibrium for Fe is well satisfied in late F-K dwarfs of any metallicity}. We
must stress, however, that while our gravities determined from masses and
luminosities are very robust (expected errors are mainly due to the adopted
temperature scale, and are smaller than 0.04 dex), our results about the
goodness of the ionization equilibrium for Fe directly depends on the adopted
model atmospheres and temperature scale, as well as on details of the
abundance analysis procedure such as the adopted oscillator strengths, for
instance. These issues will be addressed in more detail in the remaining part
of this section. 

We have 9 stars in common with Nissen et al. (1997). Our gravities average
only $0.03\pm 0.01$~dex (9 stars; r.m.s. of 0.02 dex) larger than those of
Nissen et al. (1997). This (small) systematic offset is entirely due to our
higher temperature scale, (by $119\pm 16$~K, r.m.s. scatter of 48 K).

Fuhrmann et al. (1997) derived surface gravities from the wings of strong,
pressure broadened Mg I lines. We have three stars in common with them,
(HD19445=HIP 14594, HD194598=HIP 100792, and HD201891=HIP 104659). The
Fuhrmann et al. (1997) gravities for these stars are lower than those we
derive from luminosities and masses by only $0.10\pm 0.02$\ (r.m.s.=0.04 dex),
on the whole supporting the weight Fuhrmann et al. attribute to this gravity
indicator. (The temperatures adopted by Fuhrmann et al. are very similar to
ours, differences being on average only $3\pm 11$~K, r.m.s.=19~K). With these
gravities, Fuhrmann et al. (1997) find that abundances from Fe I lines are
significantly lower than those given by Fe II lines, and suggest that some
overionization of Fe occurs in the atmospheres of solar-type stars, although
they do not rule out other possible explanations of this discrepancy. This
result is at odds with our finding. For these three stars our abundances from
Fe I lines are higher than those from Fe II lines by $0.10\pm 0.03$~dex
(r.m.s.=0.05 dex): that is, using the ionization equilibrium we would derive
gravities even larger (by $0.19\pm 0.05$~dex) than what we obtain from masses
and luminosities. The gravities Fuhrmann et al. derive for these three stars
from the equilibrium of ionization of Fe are on average $0.23\pm 0.07$~dex
smaller than obtained by us from masses and luminosities.

In order to understand the reasons for this large ($\sim 0.4$~dex) discrepancy
between gravities derived with the same method (LTE equilibrium of
ionization), we considered the Fe I and Fe II abundances separately. Once
corrected for the effects of the $\alpha-$elements enhancement, our Fe II
abundances are roughly identical to those derived by Fuhrmann et al.
(difference is $0.00\pm 0.03$~dex, r.m.s=0.06~dex). However, our abundances
from the Fe I lines are $0.17\pm 0.01$~dex (r.m.s.=0.02~dex) larger than those
by Fuhrmann et al. (1997). The reason for these differences is not clear. They
are not due to differences in the atmospheric parameters (either for the
programme stars and the Sun): in fact, if allowance is given for our larger
gravities (0.10 dex), smaller microturbulent velocities (0.34~km/s), and higher
solar temperature (5770~K rather 5750~K), we would expect our abundances to be
larger than those by Fuhrmann et al. by 0.07~dex and 0.09~dex for Fe~I and
Fe~II lines, respectively. Indeed, our Fe I abundances are unexpectedly larger
than those of Fuhrmann et al. by 0.10 dex, while those for Fe II are lower by
0.09 dex.

While both sets of results should be essentially differential with respect to
the Sun, the present analysis differs from that of Fuhrmann et al.  in a
number of respects: first, different model atmospheres are used, second, we
used laboratory oscillator strengths for Fe (see Clementini et al. ,1995, for
references). Our abundance analysis is differential in the sense that we
repeated it for both the programme stars and the Sun. The solar abundance is
determined from weak lines, and it is then insensitive to the adopted
collisional damping and to uncertainties in the equivalent widths due to
extended wings. Unfortunately these lines do not coincide with those used for
subdwarfs, and the accuracy of our abundances is determined by the reliability
of the $gf$-scale. Fuhrmann et al. preferred to use solar $gf$'s: the same
line list is then used for both the Sun and the programme stars. However,
since lines are much weaker in subdwarfs than in the Sun, solar $gf$'s for
those lines measurable in the subdwarf spectra are sensitive to the adopted
damping parameters and to the accuracy of continuum location (see Anstee,
O'Mara \& Ross, 1997). Unfortunately, errors induced by these uncertainties do
not cancel out for metal-poor dwarfs, because lines are weak in the spectra of
these stars  and damping is unimportant. Lacking more details about the line
parameters used by Fuhrmann et al., it is not clear which analysis should be
preferred, but it is evident that systematic differences as large as $\sim
0.1$~dex may well be present in the derived abundances.

Caution should be used when considering gravities derived from ionization
equilibrium, since results are influenced by the adopted temperature scale,
uncertainties in model atmospheres, and on the line parameters. However,
evidence for iron overionization in the photosphere of subdwarfs is weak. In
fact, both our and Fuhrmann et al's. results clearly exclude departures from
LTE affecting the abundances by a factor larger than 2. A further strong
argument against a significant Fe overionization in subdwarfs comes from the
extensive, consistent, statistical equilibrium calculations for Fe in the
stellar atmospheres over a wide range of temperatures and gravities by Gratton
et al. (1997b). These authors normalized the uncertain collisional cross
sections in order to reproduce observations for RR Lyraes. Since these are
warm, low gravity and metal-poor stars, overionization is expected to be much
larger than in late F-K dwarfs. The lower limit to collisional cross sections
given by the absence of detectable overionization in RR Lyrae spectra
(Clementini et al. 1995) implies that LTE should be a very good approximation
for the formation of Fe lines in dwarfs.

\subsection{O and $\alpha-$element abundances}

Oxygen abundances were derived from the permitted IR triplet, and include
non-LTE corrections computed for every line in each star following the precepts
of Gratton et al. (1997b). 

\begin{figure}
\vspace{12.0cm}
\includegraphics{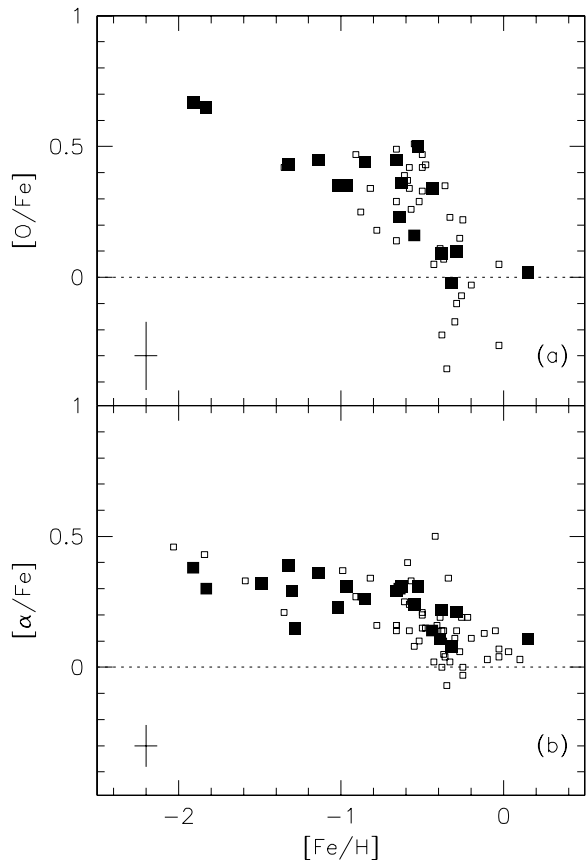}
\caption{Run of the overabundances of O (panel {\it a}) and $\alpha-$elements
(panel {\it b}) as a function of [Fe/H] for the programme subdwarfs. Filled
squares are abundances from McDonald spectra; open squares are abundances
from Asiago spectra }
\label{f:figure08} 
\end{figure}

We found that O and the other $\alpha-$elements are overabundant in all
metal-poor stars in our sample (see panels $\it a$ and $\it b$ of
Figure~\ref{f:figure08}). The average overabundances in stars with [Fe/H]$<-0.5$
are: 
$$ {\rm [O/Fe]}= +0.38\pm 0.13$$
$$ {\rm [\alpha/Fe]}= +0.26\pm 0.08,$$
where the errors are the r.m.s. scatters of the individual values around
the mean, and not the standard deviation of the sample (which is 0.02~dex in
both cases). The moderate value of the O excess derived from the IR permitted
lines is a consequence of the rather high temperature scale adopted (see also
King 1993), which directly stems from the use of the Kurucz (1993) model
atmospheres and colours. If this procedure is adopted, abundances from permitted
OI lines agree with those determined from the forbidden [OI] and the OH lines
(Carretta, Gratton \& Sneden 1997). 

\subsection{Cr and Ni abundances} 

\begin{figure}
\vspace{12.0cm}
\includegraphics{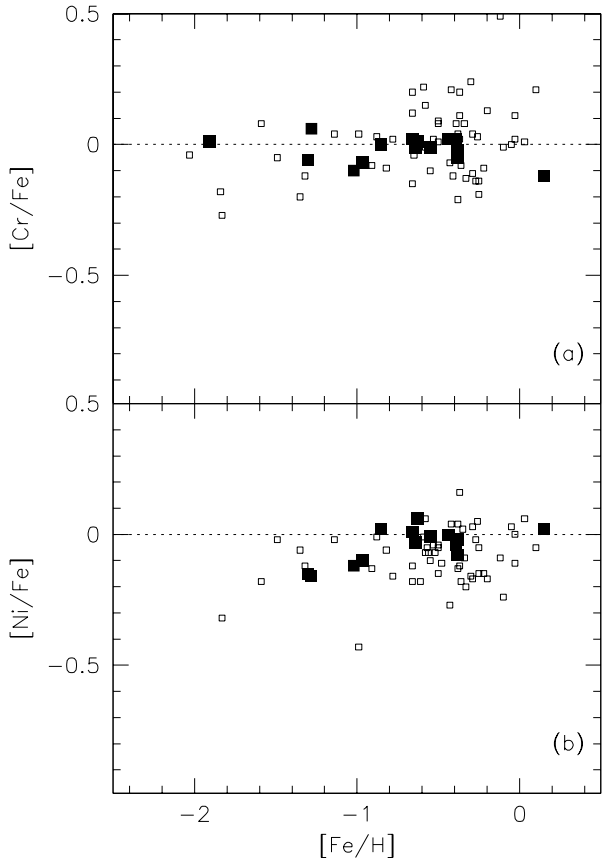}
\caption{Runs of the underabundances of Cr (panel {\it a}) and Ni (panel {\it
b}) as a function of [Fe/H] for the programme subdwarfs. Filled squares are
abundances from McDonald spectra; open squares are abundances from Asiago
spectra } 
\label{f:figure09} 
\end{figure}  

Figure~\ref{f:figure09} displays the run of the [Cr/Fe] and [Ni/Fe] abundance
ratios with [Fe/H]. Cr and Ni are very slightly deficient the most in 
metal-poor stars of our sample.

\section{Comparison with previous work}

\normalsize
\begin{table}
\begin{minipage}{160mm}
\caption{Average abundances in metal-poor stars ([Fe/H]$<-0.8$)}
\scriptsize
\begin{tabular}{lrrcrrcrc}
\hline \hline
Element&\multicolumn{3}{c}{McDonald}&\multicolumn{3}{c}{Asiago}&Others&Ref\\
\hline
${\rm [O/Fe]}$  &  7 &    0.48 & 0.13 & 5 &    0.33 & 0.12 &    0.45 & 1 \\
${\rm [Mg/Fe]}$ &  8 &    0.38 & 0.09 & ~ &    ~    & ~    &    0.38 & 1 \\
${\rm [Si/Fe]}$ &  6 &    0.32 & 0.15 & 5 &    0.22 & 0.06 &    0.30 & 2 \\
${\rm [Ca/Fe]}$ & 11 &    0.26 & 0.07 & 9 &    0.33 & 0.11 &    0.29 & 2 \\ 
${\rm [Ti/Fe]}$ & 11 &    0.28 & 0.09 & ~ &    ~    & ~    &    0.28 & 2 \\
${\rm [Cr/Fe]}$ & 12 & $-$0.05 & 0.09 & 9 & $-$0.05 & 0.10 & $-$0.04 & 2 \\
${\rm [Ni/Fe]}$ & 10 & $-$0.10 & 0.10 & 7 & $-$0.15 & 0.14 & $-$0.04 & 2 \\
\hline 
\end{tabular}
\medskip
\medskip

1. Carretta, Gratton \& Sneden 1998; 2. Gratton \& Sneden 1991

\label{t:mean}
\end{minipage}
\end{table}
A vast literature exists for some of the stars in our list. In general
agreement is quite good, once differences in the atmospheric parameters and in
the solar analysis are taken into account. The abundances presented in this
paper are on the same scale as those of Gratton et al. (1998), to which we
refer for a thorough comparison with results from other authors. The
overabundances of O and $\alpha-$ elements we find for the field subdwarfs is
also similar to the excesses found for the globular cluster giants (apart from
those stars affected by the O-Na anticorrelation, see Kraft 1994).

Table~\ref{t:mean} gives the average element-to-iron abundance ratios for
metal-poor stars ([Fe/H]$<-0.8$) obtained in this paper, along with the number
of stars and the r.m.s scatter of individual stars around the mean value.
Averages are computed separately for McDonald and Asiago spectra. For
comparison, we also give the same values derived from the analysis of Carretta,
Gratton \& Sneden (1998: O and Mg) and Gratton \& Sneden (1991: all other
elements). These authors use an abundance analysis technique similar to that
described in this paper, but different observational material. The agreement
is excellent, in particular for the McDonald spectra.

\section{Calibration of the photometric abundances}

Once combined with the abundances obtained by Gratton, Carretta \& Castelli
(1997), the sample of late F to early K-type- field stars with internally
homogenous and accurate high dispersion abundances includes nearly 400 stars.
Although the number of entries in this database may seem quite large, the vast
majority of stars of these spectral classes (the most useful in galactic
evolution studies, and those to be used to derive ages for globular clusters
via main-sequence fitting techniques) present in the HIPPARCOS catalogue still
lack of accurate metal abundances. However, Schuster \& Nissen (1989) have
shown that fairly accurate metal abundances for late F to early K-type can be
obtained using the Str\"omgren $uvby$\ photometry, which is available for a
considerable fraction of the HIPPARCOS stars. Furthermore, the extensive
binary search by Carney et al. (1994) has provided a large number of metal
abundances derived from an empirical calibration of the cross correlation dips
for metal-poor dwarfs. Finally, metal abundances from an index based on the
strength of the Ca~II K line in low dispersion spectra have been obtained by
Ryan \& Norris (1991). While the small scatter of the correlations with our
abundances (see below) is testimonial to the efforts made by these authors to
accurately calibrate their indices, their metallicity scales were at the mercy
of a heterogenous collection of literature studies based on high dispersion
analysis. In most cases these calibrating abundances were derived using model
atmospheres different from Kurucz (1993), and unable to provide solar
abundances in agreement with e.g. the meteoritic value. It is difficult to
find out what solar abundance was used in these abundance systems.

\begin{figure}
\vspace{13cm}
\includegraphics{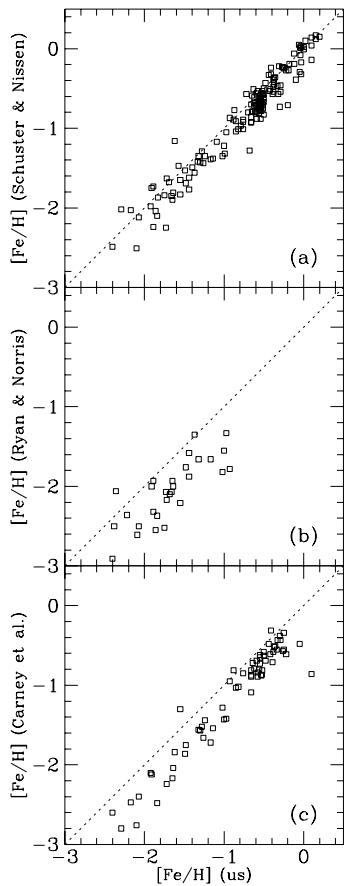}
\caption{ Comparison between the abundances obtained from high dispersion
spectra (present analysis or Gratton, Carretta \& Castelli, 1997), and those provided by the
original calibration of Schuster \& Nissen (1989, panel {\it a}), Ryan \& Norris
(1991, panel {\it b}), and Carney et al. (1994, panel {\it c}) } 
\label{f:figure10} 
\end{figure}

It is worthwhile to recalibrate these abundance scales. The Schuster \& Nissen
(1989) and Carney et al. (1994) abundances correlate very well with our high
dispersion results. Schuster \& Nissen's only differs by a zero point offset
(see panel {\it a} of Figure~\ref{f:figure10}); the mean difference is:
$${\rm [Fe/H]}_{\rm us}={\rm [Fe/H]}_{\rm SN} + (0.102\pm 0.012),$$
based on 152 stars (the r.m.s. scatter for a single star is 0.151~dex). Note
that here we considered all the stars having high dispersion abundances from
Gratton, Carretta \& Castelli (1997) and the present work, and for all these stars we derived
abundances following the precepts of Schuster \& Nissen (1989). 

In the case of Carney et al. (1994: panel {\it c} of Figure~\ref{f:figure10}), a
small linear term is also required. The best fit line (based on 66 stars) is: 
$${\rm [Fe/H]}_{\rm us}=(0.935\pm 0.032){\rm [Fe/H]}_{\rm Carney~et~al.} + 
(0.181\pm 0.173),$$
where the error on the intercept is the r.m.s. scatter around the best fit
line. 

Finally, the scatter is somewhat larger for the abundances determined by Ryan \&
Norris (1991: panel {\it b} of Figure~\ref{f:figure10}). Also, the range where these
abundances are available is quite restricted ([Fe/H]$<-1$), because the used
index saturate for more metal-rich stars. Hence, only the offset can be
determined. The best calibration we get (excluding one star, G059-27, which
gives discrepant results) is: 
$${\rm [Fe/H]}_{\rm us}={\rm [Fe/H]}_{\rm Ryan \& Norris} + (0.40\pm 0.04),$$ 
with a r.m.s. scatter of 0.23 dex for individual stars. 


The offsets between the high dispersion abundances and those provided by the
above metallicity indicators are mainly due to different assumptions about the solar
abundances in the high dispersion analyses originally used in the calibrations.
In most cases, the stellar abundances were derived using the MARCS grid of
model atmospheres (Gustafsson et al. 1975), while the solar abundances obtained
with the empirical model by Holweger \& M\"uller (1974) were used. Since the
MARCS solar model is much cooler in the line forming region than the Holweger
\& M\"uller's one, there is an offset of $\sim 0.15$\ dex between solar and
stellar abundances. This offset is canceled out when consistent model
atmospheres are used for the Sun and the stars (e.g. the solar scaled
atmospheres used sometimes, or model atmospheres extracted from the same grid).
Given the heterogenous nature of the calibrating samples, it is not possible to
go further in detail. 

Once corrected to place them on our scale, errors (derived from the r.m.s.
scatter of differences with our estimates) are 0.15~dex for abundances from the
Str\"omgren photometry and 0.18~dex for those derived from Carney et al.
(1994). For the stars having both (independent) estimates (a large fraction of
the over one thousands metal-poor dwarfs in the full HIPPARCOS catalogue)
errors are as low as 0.12~dex. 

\section{Discussion and Summary}

We have collected literature photometric data for a sample of 99 dwarfs with
parallaxes measured by the Hipparcos satellite, and high resolution spectra
have been obtained for 66 of them. The photometric data were selected through
a careful revision of the literature data, and were compared and implemented
with the Hipparcos/Tycho photometric measurements in order to get a
homogeneous, and accurate photometric data-base that includes Johnson $V$,
$B-V$, $V-K$, Cousins $V-I$ and Str\"omgren $b-y$'s. Typical accuracy are
0.014, 0.011, 0.016 and 0.022~mag in $V$, $B-V$, $V-K$ and $V-I$,
respectively.

The spectroscopic data set consists of high dispersion ($15\,000<R<60\,000$),
high $S/N$\ ($>200$) spectra obtained at the Asiago and McDonald Observatories.
They were used to measure radial velocities and to derive high accuracy abundances of Fe, O, and the
$\alpha-$elements Mg, Si, Ca, and Ti for the programme stars, according to a
procedure totally consistent with that used in Gratton, Carretta \& Castelli
(1997, $\sim 300$~field stars) and Carretta \& Gratton (1997), (giants in 24
globular clusters). 

This large and homogeneous photometric and spectroscopic data base has been
used to derive accurate ages of galactic globular clusters (Gratton et al.,
1997a). 

\bigskip\bigskip\noindent
ACKNOWLEDGEMENTS

The Hipparcos data used in the present analysis were the result of the FAST
proposal n. 022; we are grateful to P.L. Bernacca for allowing us to have
access to them and for continuous help in the related procedures. We wish to
thank G. Cutispoto for his collaboration in the data acquisition. E.Carretta
gratefully acknowledges the support by the Consiglio Nazionale delle Ricerche.
The financial support of the {\it Agenzia Spaziale Italiana} (ASI) is also
gratefully acknowledged. C. Sneden was supported by NSF grants AST-9315068
and AST-9618364.
This research has made use of the SIMBAD data-base,
operated at CDS, Strasbourg, France.


\begin{thebibliography}{}
\bibitem{} Alonso, A., Arribas, S., Martinez-Roger, C. 1994, A\&AS, 107, 365
\bibitem{} Arribas, S. \& Martinez-Roger, C. 1987, A\&AS, 70, 303
\bibitem{} Anstee, S.D., O'Mara, B.J., \& Ross, J.E. 1997, MNRAS, 284, 202
\bibitem{} Anthony-Twarog, B. J., Twarog, B. A. 1987, AJ, 94, 1222
\bibitem{} Bertelli, G., Bressan, A., Chiosi, C., Fagotto, F., \& Nasi, 
           E. 1994, A\&AS, 106,275 
\bibitem{} Bessel, M.S. 1979, PASP, 91, 589
\bibitem{} Bessel, M.S. 1990, A\&AS, 83, 357
\bibitem{} Bikmaev, I.F., Bobritskij, S.S., El'kin, V.G., Lyashko, D.A.,
             Mashonkina, L.I., \& Sakhibullin, N.A. 1990, in IAU Symp. 145,
             Evolution of Stars: the Photospheric Abundance Connection, G.
             Michaud ed.
\bibitem{} Carney, B. W. 1978, AJ, 83, 1087
\bibitem{} Carney, B. W. 1983a, AJ, 88, 610
\bibitem{} Carney, B. W. 1983b, AJ, 88, 623
\bibitem{} Carney, B. W. \& Latham, D. W. 1987, AJ, 93,116
\bibitem{} Carney, B. W. Latham, D. W., Laird, J. B., \& Aguilar, L. A. 
           1994, AJ, 107, 2240
\bibitem{} Carney, B.W., Wright, J.S., Sneden, C., Laird, J.B., Aguilar, L.A. 
           1997,  AJ, 114, 363
\bibitem{} Carretta, E., \&  Gratton, R. G. 1997, A\&AS, 121, 95
\bibitem{} Carretta, E., Gratton, R. G., \& Sneden, C. 1997, submitted to A\&AS
\bibitem{} Carretta, E., Gratton, R. G., \& Sneden, C. 1998, in preparation           
\bibitem{} Clementini, G., Carretta, E., Gratton, R. G., Merighi, R.,
           Mould, J. R., \& McCarthy, J. K. 1995, AJ, 110, 2319
\bibitem{} Cousins, A. W. J. 1976a, Mem. R.A.S., 81, 25
\bibitem{} Cousins, A. W. J. 1976b, M.N.A.S. So. Africa, 37, 62
\bibitem{} Cutispoto, G. 1991a, A\&AS, 89, 435
\bibitem{} Cutispoto, G. 1991b, A\&AS, 111, 507
\bibitem{} Edvardsson, B., Andersen, J., Gustafsson, B., Lambert, D.L., 
           Nissen, P.E., \& Tomkin, J. 1993, A\&A, 275, 101
\bibitem{} Eggen, O. J. 1955, AJ, 60, 65
\bibitem{} Eggen, O. J. 1956, AJ, 61, 462
\bibitem{} Eggen, O. J. 1968a, ApJS, 16, 97
\bibitem{} Eggen, O. J. 1968b, ApJS, 153, 195
\bibitem{} Eggen, O. J. 1972, ApJ, 175, 787
\bibitem{} Eggen, O. J. 1976, PASP, 88, 732
\bibitem{} Eggen, O. J. 1978, ApJS, 37, 251
\bibitem{} Eggen, O. J. 1979, ApJ, 229, 158
\bibitem{} Eggen, O. J. 1987a, AJ, 93, 393
\bibitem{} Eggen, O. J. 1987b, AJ, 92, 379
\bibitem{} Feltzing, S., \& Gustafsson, B. 1998, A\&AS 129, 237
\bibitem{} Fuhrmann, K., Pfeiffer, M., Frank, C., Reetz, J., Gehren, T. 1997,
           A\&A, 323, 909 
\bibitem{} Glass, I. S. 1974, MNSSA, 33, 53
\bibitem{} Gratton, R. G., Carretta, E., \& Castelli, F. 1997, A\&A, 314, 191
\bibitem{} Gratton, R. G., Carretta, E., Gustafsson, B., \& Eriksson, K. 1997b, 
           submitted to A\&A
\bibitem{} Gratton, R. G., Fusi Pecci, F., Carretta, E., Clementini, G., 
           Corsi, C. E., Lattanzi, M. 1997a, ApJ, 491, 749 
\bibitem{} Gratton, R. G., Carretta, E., Matteucci, F., \& Sneden, C. 1998 in 
           preparation
\bibitem{} Gratton, R. G., \& Sneden, C. 1991, A\&A, 241, 501  
\bibitem{} Grossmann, V. et al. 1995, A\&A, 304, 110
\bibitem{} Gustafsson, B., Bell, R.A., Eriksson, K., \& Nordlund, A. 1975,
             A\&A, 42, 407
\bibitem{} Holweger, H., \& M\"uller, E.A. 1974, Solar Phys., 39,19
\bibitem{} Johnson, H. L., Mitchell, R. I., Iriarte, B., \& Wi\'sniewski, W. 
             Z. 1966, Comm. Lunar Planet. Lab., 4, 99
\bibitem{} King, J. R. 1993, AJ, 106, 1206
\bibitem{} King, J.R. 1997, AJ, 113, 2302
\bibitem{} Kraft, R. P. 1994, PASP, 106, 553
\bibitem{} Kron, G. E., White, H. S., \& Gascoigne, S. C. B. 1953, ApJ,
             118, 502
\bibitem{} Kroupa, P., Tout, C.A., \& Gilmore, G. 1993, MNRAS, 262, 545
\bibitem{} Kurucz, R. L. 1993, CD-ROM 13 and CD-ROM 18
\bibitem{} Laird, J.B. 1985, ApJS, 57, 389
\bibitem{} Laird, J.B., Carney, B. W., \& Latham, D. W. 1988, A.J., 95, 1843
\bibitem{} Lutz, T. E., Kelker, D. H. 1973, PASP, 85, 573
\bibitem{} Magain, P., \& Zhao, G. 1996, A\&A, 305, 245 
\bibitem{} Nissen, P.E., H\/og, E., Schuster, W.J., 1997, in ESA Symposium 
           "Hipparcos-Venice '97", ed. B. Battrick, ESA SP-402, Noordwijk, 
           p. 225
\bibitem{} Olsen, E.H. 1983, A\&AS, 54, 55
\bibitem{} Olsen, E.H. 1984, A\&AS, 57, 443
\bibitem{} Olsen, E.H. 1994a, A\&AS, 104, 429
\bibitem{} Olsen, E.H. 1994b, A\&AS, 106, 257
\bibitem{} Perryman, M.A.C., 1989 ed. The HIPPARCOS Mission, Pre-Launch
             Status, ESA SP1111, 515 pag.
\bibitem{} Peterson. R.C., Willmarth, D.W., Carney, B.W., \& Chaffee, F.H.Jr,
           1980, ApJ 239, 928
\bibitem{} Pilachowski, C., Sneden, C., \& Wallerstein, G., 1983, ApJS 52,241
\bibitem{} Ryan, S.G., Norris, J.E., 1991, AJ, 101, 1835
\bibitem{} Sandage, A., \& Kowal, C. 1986, AJ, 91,1140
\bibitem{} Schuster, W. J., \& Nissen, P. E. 1988, A\&A, 73, 225
\bibitem{} Schuster, W. J., \& Nissen, P. E. 1989, A\&A, 221, 65
\bibitem{} Schuster, W. J., Parrao, L., \& Contreras Martinez, M. E.
             1993, A\&AS, 97, 951
\bibitem{} Tomkin, J., Lemke, M., Lambert, D.L., \& Sneden, C. 1992, AJ, 104, 
             1568
\bibitem{} Twarog, B. A., \& Anthony-Twarog, B. J.  1995, AJ 109, 2828 
\bibitem{} Upgren, A. R. 1974, PASP 86, 294
\bibitem{} Wallerstein, G., Greenstein, J.L., Parker, R., Helfer, H.L., \&
           Aller, L.H., 1963, ApJ 137, 280
\bibitem{} Wheeler, J.C., Sneden, C., \& Truran, J.W. 1989, ARAA, 27, 279
\bibitem{} Weis, E. W. 1996, AJ, 112, 2300
\end{thebibliography}
\end{document}